\documentstyle[aps,preprint,epsfig,eqsecnum]{revtex}
\tighten
\begin{document}
\preprint{\begin{minipage}{33mm}{\footnotesize
TU-741\\%
KYUSHU-HET-81\\%
hep-ph/0504036\\[2mm]}
\end{minipage}}
\title{Bottom-Up Approach to Moduli Dynamics\\[0.7mm]%
in Heavy Gravitino Scenario :\\[1mm]%
Superpotential, Soft Terms and Sparticle Mass Spectrum}
\author{Motoi Endo$^1$, Masahiro Yamaguchi$^1$, and Koichi Yoshioka$^2$}
\address{\medskip%
$^1$Department of Physics, Tohoku University, Sendai, 980-8578, Japan}
\address{%
$^2$Department of Physics, Kyushu University, Fukuoka, 812-8581, Japan}
\maketitle

\begin{abstract}%
The physics of moduli fields is examined in the scenario where the
gravitino is relatively heavy with mass of order 10~TeV, which is
favored in view of the severe gravitino problem. The form of the
moduli superpotential is shown to be determined, if one imposes a 
phenomenological requirement that no physical CP phase arise in
gaugino masses from conformal anomaly mediation. This bottom-up
approach allows only two types of superpotential, each of which can
have its origins in a fundamental underlying theory such as
superstring. One superpotential is the sum of an exponential and a 
constant, which is identical to that obtained by Kachru et 
al.\ (KKLT), and the other is the racetrack superpotential with two
exponentials. The general form of soft supersymmetry breaking masses
is derived, and the pattern of the superparticle mass spectrum in the
minimal supersymmetric standard model is discussed with the KKLT-type
superpotential. It is shown that the moduli mediation and the anomaly
mediation make comparable contributions to the soft masses. At the 
weak scale, the gaugino masses are rather degenerate compared to the
minimal supergravity, which bring characteristic features on the
superparticle masses. In particular, the lightest neutralino, which
often constitutes the lightest superparticle and thus a dark matter
candidate, is a considerable admixture of gauginos and higgsinos. We
also find a small mass hierarchy among the moduli, gravitino, and
superpartners of the standard-model fields. Cosmological implications
of the scenario are briefly described.
\end{abstract}

\renewcommand{\thefootnote}{\fnsymbol{footnote}}
\footnote[0]{{\footnotesize Preliminary results have been presented 
at YITP workshop {\it Progress in Particle Physics}, Kyoto, Japan, 
June 29--July 2, 2004.}}
\renewcommand{\thefootnote}{\arabic{footnote}}
\setcounter{footnote}{0}

\newpage

%%%%%%%%%%%%%%%%%%%%%%%%%%%%%%%%%%%%%%%%%%%%%%%%%%%%%%%%%%%%%%%%%%%%%%
\section{Introduction}

Low-energy supersymmetry provides us with one of the most attractive
candidates for the fundamental theory beyond the standard
model~\cite{SUSY}. However supersymmetry must be a broken symmetry
below the electroweak scale due to the absence of any experimental
signatures. Supersymmetry breaking is expressed by soft terms which do
not reintroduce quadratic divergences~\cite{soft} and spoil the
Planck/weak scale hierarchy~\cite{hierarchy}. These soft breaking
terms consist of gaugino masses, scalar masses, and scalar trilinear
couplings. Generic forms of soft breaking terms lead to new
contributions and sometimes disastrous phenomenological effects to
flavor-changing rare processes~\cite{FCNC} as well as to CP
violation~\cite{CP}. To satisfy the experimental constraints and to
make supersymmetric models viable, supersymmetry-breaking terms are
forced to have special properties, which could be realized in various
ways proposed in the literature.

In the context of supergravity and superstring theory, there exists
the gravitino, namely, the supersymmetric partner of the graviton. It
interacts with all particle species through tiny Planck-suppressed
couplings and may be hard to be observed in collider experiments (see, 
however,~\cite{collider-g}.) \ On the other hand, the gravitino is too
easily ``detected'' in a cosmological sense and, in turn, that leads to
severe constraints on its property in almost all supersymmetric
theories~\cite{g-problem}. This gravitino problem exists even in the
inflationary Universe. If the gravitino is stable, its relic abundance
has to be small not to contribute too much to the energy density of
the Universe~\cite{light-g}. This can be achieved either if the
gravitino mass is sufficiently light~\cite{light-g}, or if the
reheating temperature of the inflation is low enough to suppress the
regeneration of gravitinos in the thermal bath~\cite{lowtemp}. For
an unstable gravitino, the most severe constraint comes from the
big-bang nucleosynthesis. If gravitinos decay with electromagnetic
and/or hadronic showers during or after the nucleosynthesis epoch, the
decay products would spoil the successful predictions of the big-bang
nucleosynthesis by destroying synthesized light
elements~\cite{g-problem,BBNwX_old,photodis_new,Kohri,hadronic}. This
argument puts a severe constraint on the gravitino abundance, which
may be suppressed by invoking the inflation with sufficiently
low reheating temperature. The constraint becomes relaxed and
eventually disappears as the gravitino mass increases and it decays
earlier. A heavy gravitino with mass of order 10~TeV can therefore
greatly ameliorate the problem.

Supersymmetric theories also include particles with masses around 
the electroweak scale and with suppressed couplings to ordinary
matter. They are generally called moduli fields, the existence of
which may be related to the degeneracy of vacua parameterized by
symmetries in high-energy theory. Therefore the moduli potential is
flat in perturbation theory and only lifted by non-perturbative
effects or supersymmetry breaking, which induces masses of moduli
fields around the supersymmetry-breaking scale. It is known that the
existence of moduli leads to serious cosmological
difficulties~\cite{m-problem} with their suppressed couplings and
resultant long lifetime. Moduli fields start to oscillate when the
expansion rate of the Universe becomes smaller than moduli masses. The
coherent oscillations decay into ordinary particles after the
nucleosynthesis occurred for moduli masses around the electroweak
scale. With a natural value of the initial amplitude of moduli,
induced huge entropy destroys the great success of the big-bang
nucleosynthesis as a heavy gravitino does. This is the cosmological
moduli problem. A simple and attractive way to evade the problem is to
make all moduli sufficiently heavy so that the moduli decay before the
nucleosynthesis~\cite{heavy-m,MYY,RT,KYY1,KYY2}. The masses of moduli
fields might naively be expected to be of the same order of other
supersymmetry-breaking mass parameters, especially a similar order to
the gravitino mass as in the hidden sector models. However there could
be some hierarchy between them, crucially depending on the form of
moduli potential which originates from supersymmetry-breaking
dynamics. In these ways, cosmological arguments may be strong enough
to discriminate dynamics of supersymmetry breaking.

In this paper, we consider the physics of moduli fields in the heavy
gravitino scenario, i.e.\ with supersymmetry-breaking scale higher
than the electroweak scale. This may be rather a common situation in
supersymmetric theories, including supergravity and superstring
theory. The only few exceptions include, e.g.\ supersymmetry breaking
with strongly-coupled gauge dynamics in a low-energy
regime~\cite{GM}. The heavy gravitino generally implies that moduli
fields also have their masses around the gravitino mass scale, and
therefore the cosmological gravitino/moduli problems may be solved by
making these fields sufficiently heavy. As mentioned above, the masses
of moduli fields crucially depend on their potential form, which is
supposed to arise from (non-perturbative) dynamics of supersymmetry
breaking. There have been various proposals for potential forms
derived from specific non-trivial dynamics in high-energy theory. A
well-known example is for the dilaton modulus field in string-inspired
theories, which modulus is stabilized by (multiple) gaugino
condensations~\cite{gaugino,racetrack,gaugino2}, non-perturbative
corrections to K\"ahler terms~\cite{Kahler-corr}, and so on. Recent
development of the flux compactifications in type IIB superstring
theory reveals another possibility, in which the dilaton is fixed in
the presence of fluxes, and the moduli stabilization can be achieved
by some additional non-perturbative effects~\cite{flux,KKLT}. Patterns
of supersymmetry breaking by minimizing scalar potential have been
discussed in Ref.~\cite{pattern}.

We pursue, in a way, an opposite approach in that the possible form
of moduli potential is highly constrained from a phenomenological
viewpoint without referring to any specific dynamics at high-energy
regime. For that purpose, it is a key ingredient to take into account
the supersymmetry-breaking effects associated with the superconformal
anomaly~\cite{AM}. The soft mass parameters from the anomaly mediation
are loop suppressed (because of being related to quantum anomalies)
relative to the gravitino mass. It should be noted that this
anomaly-mediated contribution ubiquitously exists in all scenarios of
supersymmetry breaking. Such unavoidable anomaly effects cannot be
neglected in the heavy gravitino models and may provide significant
corrections to observable quantities measured at high
accuracy~\cite{EYY}. Among them, the violation of CP invariance is one
of the most notable quantities, including the electric dipole moments
of nucleons, atoms, and charged leptons. The experimental results
indicate that complex phases of soft mass parameters are tightly
constrained so that supersymmetric contributions to CP-violating
observables do not exceed the predictions of the standard model. On
the basis of such phenomenological results, we will find in Section II
that the superpotential of moduli fields which participate in
supersymmetry breaking is uniquely determined (up to a K\"ahler
transformation) by requiring CP be automatically conserved.

It is worth mentioning that the possible form of a moduli potential is
completely fixed irrespectively of the stabilization and specific
dynamics of moduli fields. Interestingly enough, we will see in
Section III that the obtained potential, however, does stabilize the 
moduli and also cause supersymmetry breaking. Furthermore the
superpotential has its origin in ordinary field theory and superstring
theory. In particular, it has recently been 
shown~\cite{flux,KKLT} that some classes of superstring
compactification in the presence of non-trivial fluxes generate the
above-mentioned superpotential for K\"ahler moduli fields. It may be a
surprising result that two completely different perspectives, a 
top-down one from string theory and a bottom-up one from experimental
observations, derive the same and unique form of moduli potential.

It is found in Section IV that the mass spectrum in the model with the
uniquely-determined superpotential exhibits two interesting types of
small hierarchies among superparticles, gravitino, and moduli
fields. The lightest particles are gauginos and scalar partners of
quarks and leptons with masses around the electroweak or TeV
scale. These mass scales are suppressed roughly by one-loop factors
compared to the gravitino mass due to the loop-suppressed contribution
of the anomaly mediation. This first hierarchy is favorable for
solving the gravitino problem as stated above. The heaviest states are
moduli whose masses are O(10) times that of the gravitino. This second
hierarchical factor is roughly determined by the logarithm of the
ratio between the Planck and electroweak scales. That is naturally 
obtained by minimizing the uniquely determined moduli potential. Since
the moduli become heavy in the vacuum, they have suppressed
contributions to supersymmetry breaking. We find that the anomaly and
moduli mediations induce numerically similar sizes of soft mass
parameters for visible-sector superpartners, which are of the order of
the electroweak scale.  It is noted that, as mentioned above, our
analysis covers a mass spectrum derived from some superstring theory in
flux compactifications. The problem of the tachyonic slepton, which is a
serious problem in the model of pure anomaly mediation, is
ameliorated. More interestingly, our scenario predicts the lightest
neutralino with a significant composition of higgsinos as the lightest
supersymmetric particle in a wide range of parameter space. Together
with this issue, we study in Section V new cosmological implications
of our model with heavy moduli fields.

%%%%%%%%%%%%%%%%%%%%%%%%%%%%%%%%%%%%%%%%%%%%%%%%%%%%%%%%%%%%%%%%%%%%%%
\section{The Model}
%%%%%%%%%%%%%%%%%%%%%%%%%%%%%%%%%%%%%%%%%%%%%%%%%%%%%%%%%%%%%%%%%%%%%%
\subsection{Four-dimensional supergravity}

The dynamics of gravitino and moduli fields is described 
by $N=1$ four-dimensional supergravity. The gauge-invariant
Lagrangian, in particular the scalar potential, is most simply written
down with the compensator formalism of
supergravity~\cite{SCsugra}. Larger gauge symmetries make the theory
easily analyzed, and the usual Poincar\'e supergravity is
obtained by fixing redundant conformal/Poincar\'e local gauge
symmetries. Let us consider a single chiral superfield $X$, for
simplicity, but it is straightforward to include multiple superfields
in the Lagrangian. The most general supergravity Lagrangian is given by
\begin{equation}
  e^{-1}{\cal L}_c \,=\, \int\!d^4\theta\,\Phi^\dagger\Phi\,
  f(X,X^\dagger) \,+\int\!d^2\theta\,\Phi^3\,W(X) +{\rm h.c.},
  \label{L}
\end{equation}
where $e$ is the determinant of the vierbein field and the reduced
Planck mass is set to unity. We use as usual the same notation $X$ for
a chiral superfield, scalar component, and its vacuum expectation
value. $W(X)$ is the superpotential of $X$, and $f$ the 
supergravity $f$ function which is related to the K\"ahler 
potential $K$ in the Einstein frame
\begin{equation}
  f(X,X^\dagger) \,=\, -3e^{-K(X,X^\dagger)/3}.
\end{equation}
The symbols $\int\!d^4\theta$ and $\int\!d^2\theta$ mean the uses 
of $D$- and $F$-type action formulas of superconformal tensor
calculus. The chiral superfield $\Phi$ is the conformal compensator
multiplet and its value is fixed by the part of superconformal gauge
transformation such as dilatation so that $\Phi=1+F_\phi\theta^2$. In
the discussion below, we assume the flat gravitational background and
will drop $e$ from the action. The Lagrangian (\ref{L}) contains the
following terms of auxiliary $F$ components
\begin{equation}
  {\cal L}_c \,=\, fF_\phi^* F_\phi +f_XF_\phi^*F_X 
  +f_{\bar X}F_\phi F_X^\dagger +f_{X\bar X}F_X^\dagger F_X 
  +(3WF_\phi+W_XF_X +{\rm h.c.})+\cdots.
  \label{Laux}
\end{equation}
The lower indices of $f$ and $W$ denote the field derivatives. The
equations of motion are thus given by
\begin{eqnarray}
  F_X &\,=\,& -e^{K/3}K_{X\bar X}^{\,-1}(W_X+WK_X)^\dagger, 
  \label{F_X} \\
  F_\phi \,&\,=\,& e^{K/3}W^* +\frac{1}{3}K_XF_X. \label{F_phi}
\end{eqnarray}
Substituting these equations back into (\ref{Laux}), one obtains the
scalar potential of the $N=1$ four-dimensional supergravity
\begin{equation}
  V \,=\, e^{K/3}\Big[\,(W_X+WK_X)^\dagger K_{X\bar X}^{\,-1} 
  (W_X+WK_X) -3|W|^2\,\Big]
\end{equation}
in the conformal frame. The canonical potential in the Einstein frame
is easily obtained by field-dependent Weyl rescaling. Assuming that
the vacuum expectation value of the scalar potential vanishes, a
location of the potential minimum is unchanged by this Weyl rescaling,
and thus we will use the conformal frame in most of the subsequent
discussions.

As seen from the above formulation, there are two types of
supersymmetry-breaking $F$ terms. One is the chiral 
superfield $F$ component $F_X$, where $X$ participates in
supersymmetry breaking with a non-vanishing $F_X$. Another is the
compensator contribution $F_\phi$ that should be taken into account in
any supersymmetry-breaking scenario. In particular, 
since $F_\phi$ gives a gravitino mass scale, its contribution to other
superparticles is important in the heavy gravitino scenario. In the
present work, that will play a key role to determine the potential
form of moduli fields.

In the superconformal framework, the Lagrangian of vector multiplets,
especially for gaugino masses, includes the gauge kinetic function $S$,
\begin{equation}
  {\cal L}_v \,=\, \int\!d^2\theta\,
  S\Big(X,\frac{\mu}{\Lambda \Phi}\Big) W^\alpha W_\alpha +{\rm h.c.}.
  \label{Lv}
\end{equation}
At the classical level, the compensator $\Phi$ does not appear in the
gauge kinetic term as the gauge chiral superfield $W^\alpha$ has a
chiral weight $\frac{3}{2}$. It turns out that the dependence 
of $\Phi$ comes out radiatively through the ultraviolet 
cutoff $\Lambda$ ($\mu$ is the renormalization scale). In this paper,
we take for simplicity $S\propto X$ by holomorphic field
redefinition. Note however that this choice does not necessarily 
means $X$ is a dilaton in the four-dimensional supergravity, whose
vacuum expectation value determines the gauge coupling 
constant $g$. One may instead suppose, 
e.g.\ $S=\frac{1}{4g^2}+X$, but there are only quantitative difference
between these two forms of gauge kinetic function. In particular, the
following analysis of potential forms of moduli is completely
unchanged.

A gaugino receives two types of supersymmetry-breaking mass from the
Lagrangian (\ref{Lv}). As mentioned above, the effect of the
compensator multiplet appears only through the renormalization and
induces gaugino masses which are determined by super Weyl anomaly. The
total gaugino mass $M_\lambda$ in the canonically normalized basis is
found
\begin{equation}
  M_\lambda \,=\, \frac{F_X}{2X_R} + \frac{\beta_g}{g}F_\phi,
  \label{gauginomass}
\end{equation}
where $\beta_g$ is the gauge beta function
($\beta_g=\frac{\partial g}{\partial\ln\mu}$). We have defined the
real part of $X$ as $X_R\equiv{\rm Re}\,X$. Gaugino masses given above
are evaluated at a high-energy scale. Here we simply neglect threshold
corrections from high-energy theory. When only the moduli- and
anomaly-mediated contributions are taken into account, higher-order
corrections and threshold effects do not disturb the discussion about
complex phases of $M_\lambda$. These corrections just modify the mass
spectrum slightly.

%%%%%%%%%%%%%%%%%%%%%%%%%%%%%%%%%%%%%%%%%%%%%%%%%%%%%%%%%%%%%%%%%%%%%%
\subsection{Gaugino mass phases and moduli potential}

We would like to investigate the dynamics of moduli fields. As can be
seen in (\ref{L}), that is determined by two functions of the moduli
field $X$; K\"ahler potential $K(X,X^\dagger)$ and 
superpotential $W(X)$. We first note that the assumption that $X$ is a
modulus gives a constraint on its K\"ahler potential. The existence of
moduli fields may be related to the degeneracy of vacua which are
parameterized by symmetries of theory. Then imaginary parts of moduli
fields often transform non-linearly under these flavor symmetries
which dictate the moduli K\"ahler potential in low-energy effective
theory as
\begin{equation}
  K(X,X^\dagger) \,=\, K(X+X^\dagger)
  \label{kahler}
\end{equation}
with a real function $K$. We assume that this shift symmetry would be
broken only to an extent that it does not induce any observable
effects. On the other hand, the moduli superpotential $W(X)$ is not 
allowed by the shift symmetry as it should be, and may be generated due
to symmetry-breaking effects in effective theory. Thus, moduli
superpotential does not generally have any restrictions and its
promising forms have been discussed in model-dependent ways so as to
stabilize moduli expectation values, to have moduli supersymmetry
breaking, etc. We will show below that, in the heavy gravitino
scenario, a possible form of moduli superpotential can be determined
model-independently and also without taking any particular assumption
of the moduli K\"ahler potential (\ref{kahler}).

To this end, it is important to notice that there are two types of
supersymmetry-breaking masses of gauginos (\ref{gauginomass}). In the
heavy gravitino scenario, these two contributions may be comparable in
size to each other. Therefore their relative phase value gives rise to
sizable complex phases of gaugino masses. If there is only one gaugino
in the theory, its mass can be made real 
with $U(1)_R$ rotation. However, realistic models such as
supersymmetric standard models may contain multiple gauge factors. In
this case, phases of gaugino masses are generally not aligned because
the anomaly mediated contributions ($F_\phi$ terms) are proportional
to gauge beta functions which are diverse from each
other. Consequently, all but one gaugino mass phase cannot be rotated
away from the Lagrangian and remain physical observables.\footnote{In
four-dimensional theory, $U(1)_R$ is the only global symmetry under
which phases of gaugino masses are shifted. This is because of the 
theorem~\cite{WW} saying that massless vector fields cannot couple to
any Lorentz-covariant currents. Thus gauginos can only couple to the
symmetries which do not commute with supersymmetry.} It is easily
found that the situation is not improved even if the standard gauge
groups are embedded into a unified gauge group in a high-energy
regime. This is because anomaly mediated contributions are determined
only by low-energy quantities and decouple from high-energy
physics. In other words, threshold effects at the unification scale
split the complex phases of low-energy gaugino masses. It might be an
interesting possibility that non-vanishing phases of gaugino masses
are detectable in future particle experiments~\cite{EYY}. But, in that
case, other parameters should be carefully chosen so as to satisfy
various severe constraints which come from approximate CP
conservations observed in nature.

In this work, we pursue a more natural way that each gaugino mass has
no more tiny phase; namely, all gaugino mass phases are aligned, which
are rotated away by a single $R$ rotation. It is clear
from (\ref{gauginomass}) that this is achieved by realizing a tiny
relative phase between the two supersymmetry-breaking $F$ terms. The
ratio is simply written as
\begin{equation}
  \frac{F_\phi}{F_X} \,=\, -[\ln(fG_{\bar X})]_X,
\end{equation}
where $G$ is the supergravity K\"ahler 
potential: $G\equiv K+\ln|W|^2$. But it is suitable for practical
purpose that the ratio is rewritten with $K$ and $W$;
\begin{equation}
  \frac{F_\phi}{F_X} \,=\, \frac{K_X}{3}
  -\frac{K_{X\bar X}}{K_{\bar X}+\big(\frac{W_X}{W}\big)^\dagger}.
  \label{ratio}
\end{equation}
Since we now consider $X$ as a moduli field, any derivatives of the
K\"ahler potential $K$ are real valued, irrespectively of its detailed
form. When this ratio of two $F$ terms is real, all gauginos receive
supersymmetry-breaking masses with aligned complex phases and they can
be made real. We find from Eq.~(\ref{ratio}) that this is the case if
\begin{equation}
  \frac{W_X}{W} \,=\, {\rm real}.
  \label{cpcond}
\end{equation}

The condition (\ref{cpcond}) must be satisfied in the vacuum of the
supergravity scalar potential. With the moduli K\"ahler potential
(\ref{kahler}), it is easy to first minimize the scalar potential with
respect to the imaginary part of $X$ ($X_I\equiv{\rm Im}\,X$);
\begin{equation}
  0 \,=\, \frac{\partial\,V}{\partial X_I} \,=\, 
  -ie^{K/3} K_{X\bar X}^{\,-1}\Big[\,(WK_X +W_X)(W_XK_X+W_{XX})^\dagger
  +3K_{X\bar X}W_XW^*-{\rm h.c.}\,\Big].
\end{equation}
This leads to a condition
\begin{equation}
  \bigg(K_XK_{\bar X}-3K_{X\bar X}
  +K_{\bar X}\frac{W_{XX}}{W_X}\bigg)|W|^2
  \frac{W_X}{W} +|W_X|^2\frac{W_{XX}}{W_X} \;=\; {\rm real}.
\end{equation}
It is found that when gaugino mass phases are small, namely, the
condition (\ref{cpcond}) is satisfied, the following equation:
\begin{equation}
  \frac{W_{XX}}{W_X} \;=\; {\rm real}
  \label{WXXWX}
\end{equation}
is also satisfied in the vacuum as long as the moduli field causes
supersymmetry breaking ($F_X\neq0$). The reverse is also true that if
one has superpotential satisfying $W_{XX}/W_X$ = real, 
then $W_X/W$ takes a real value and the condition for aligned gaugino
masses is certainly realized at the extrema of supergravity
potential. It is important that the equation (\ref{WXXWX}) is
integrable. That is, since the superpotential $W(X)$ is a holomorphic
function, $W_{XX}/W_X$ should be a real constant with respect 
to $X$. (It may be rather unlikely that the reality condition
is satisfied with real coupling constants and/or vacuum expectation
values. That might be realized only in a special circumstance where an
original theory contains real parameters and would also require
additional requirements.) \ We thus find the most general moduli
superpotential leading to aligned gaugino masses
\begin{equation}
  W(X) \,=\, ae^{-bX}+c,
  \label{spot1}
\end{equation}
where $a$, $b$, and $c$ are $X$-independent factors and, in
particular, $b$ must be real. The negative sign in front of $b$ is
just a convention. We have found that, with the moduli superpotential
(\ref{spot1}), all gaugino masses can be made real in the extrema of
supergravity scalar potential and do not induce large CP-violating
operators. It should be noted that the result has been derived without 
specifying a detailed form of the moduli K\"ahler potential. In the
complete analysis of scalar potential with a fixed form of $K$, one
should check whether it is a (at least local) minimum of the
potential. That will be examined in the next section.

We have so far focused only on the phases of gaugino mass
parameters. Similar arguments are also applied to other
supersymmetry-breaking parameters. Let us define, as usual, a
holomorphic soft breaking parameter as a ratio of a coupling of
holomorphic supersymmetry-breaking term to a corresponding
superpotential one. These soft-breaking parameters are unaffected by
phase redefinitions of chiral superfields. We will explicitly show in
the next section that, if the condition (\ref{cpcond}) for aligned
gaugino masses is satisfied, all complex phases of holomorphic soft
parameters are aligned to that of gaugino masses and become real once
gaugino masses are rotated to be real by $R$ rotation.

Our argument above is based on the supergravity scalar potential which
is controlled by two functions $K(X,X^\dagger)$ and $W(X)$. It is
known that the supergravity potential is invariant under K\"ahler
transformation involving these two functions:
$K\to K+\Gamma(X)+\bar\Gamma(\bar X)$ and $W\to e^{-\Gamma(X)}W$
with an arbitrary function $\Gamma(X)$. Noting that the K\"ahler
potential depends on the moduli in the form of (\ref{kahler}), the
only possible transformation is given by $\Gamma(X)=b'X$ with a
real factor $b'$. After such transformation, we have the superpotential
\begin{equation}
  W(X) \,=\, a_1e^{-b_1X}+a_2e^{-b_2X},
  \label{spot2}
\end{equation}
where $a_{1,2}$ and $b_{1,2}$ are complex and real $X$-independent
factors, respectively. This is another form of moduli superpotential
which satisfies our requirement that CP invariance is not broken by
the moduli physics. That is, complex phases of soft
supersymmetry-breaking parameters are aligned in the vacuum of
supergravity potential. One should note that this form of the moduli
superpotential as a solution to the supersymmetric CP problem was
recognized in Ref.~\cite{Choi}\footnote{We thank K.~Choi for drawing our
attention to this paper.}

%%%%%%%%%%%%%%%%%%%%%%%%%%%%%%%%%%%%%%%%%%%%%%%%%%%%%%%%%%%%%%%%%%%%%%
\subsection{Origins of moduli superpotential}

In the previous section we found two types of superpotential for
moduli fields. With these forms of superpotential, one obtains
phenomenologically favorable results; namely, CP-violating effects of
moduli are automatically suppressed in the vacuum. On the other hand,
more formal aspects, e.g.\ the questions of whether such potentials
are derived from some high-energy theory and which of these two forms
is relevant, may depend on the property of moduli fields and its
natural form of K\"ahler potentials.

For some specific moduli fields, the superpotential forms
(\ref{spot1}) and (\ref{spot2}) have been discussed in the 
literature. Among these is that $X$ is the dilaton superfield
in four-dimensional effective theory of superstring. In this case, an
exponential factor $e^{-bX}$ emerges through gaugino
condensation~\cite{gaugino,racetrack,gaugino2} where the 
constant $b$ is inversely proportional to the gauge beta function and
therefore a real parameter, just as needed. A likely example of
dilaton stabilization along this line is the racetrack
mechanism~\cite{racetrack} with superpotential of the form
(\ref{spot2}). However it was found that a realistic value of gauge 
coupling is achieved only in rather complicated models, for example,
with more than two condensations. The constant $c$ term in
(\ref{spot1}) is also known to play important roles of the dilaton
stabilization and supersymmetry breaking~\cite{gaugino2,RW}. We should
note that such a type of superpotential with the constant term is
obtained in a five-dimensional model~\cite{LS} where $X$ is the radius
modulus which determines the size of the compact fifth dimension.

As we mentioned in the introduction, it is revealed in perturbative
string theory that there exist some massless moduli fields whose
presence may be in contradiction to low-energy experiments. Recently,
it has been discussed that all such moduli can be stabilized in some
string compactifications with non-trivial fluxes. As an example, let
us consider a class of type IIB orientifold compactification. The
absence of non-trivial fluxes classically leads to supersymmetric
vacua which are described by the dilaton modulus, complex 
structure (shape) moduli, and complexified K\"ahler (size) moduli. If
three-form field strengths are turned on, they give rise to a
tree-level superpotential $W_0$~\cite{GVW}. Having this superpotential
in hand, the vacuum is explored with effective four-dimensional
supergravity potential. It is found that for a generic choice of
fluxes, the minimization of potential fixes all shape moduli and the
dilaton~\cite{flux}. This scheme of stabilization sets the expectation
value of the dilaton to a weak coupling and is under reliable control.

The above superpotential $W_0$ is, however, independent of the
K\"ahler size moduli and they can take arbitrary values. One needs
some additional potential which might be generated by stringy and
quantum corrections to K\"ahler and/or superpotentials. It is
suggested by Kachru-Kallosh-Linde-Trivedi (KKLT) in~\cite{KKLT} that
the K\"ahler size moduli can be stabilized by including
non-perturbative corrections~\cite{nonpert} to the tree-level
superpotential. These corrections are known to depend
on the size moduli and be controlled at weak
coupling. Ref.~\cite{KKLT} discussed two types of such
non-perturbative effects: (i) D-brane instantons and (ii) gaugino
condensation. In the case (i), the source of superpotential is
D3-brane instantons wrapped on surfaces in the Calabi-Yau 
three-fold. The instanton-induced superpotential takes the 
form $W_{\rm np}=\sum a_ne^{-2\pi n\cdot\chi}$ where $\chi$ are the
complexified volumes, which are natural holomorphic coordinates
appearing in type IIB orientifold compactification. In the large
volume limit, the shape moduli are stabilized by the leading
contribution $W_0$. Accordingly one can use $W=W_0+W_{\rm np}$ as
effective superpotential for the K\"ahler size moduli. In the case 
of (ii), the superpotential is supposed to be generated by
non-perturbative effects in world-volume gauge theory on D7 branes. The
gauge coupling in this theory depends on the volume $V$ and thus
non-perturbative effects lead to a 
superpotential $W_{\rm np}\sim e^{-V/\beta}$. The constant $\beta$ is
real and related to the gauge beta function which is of 
order $N_c$ for $SU(N_c)$ gauge theory. For a small value of the flux
superpotential $W_0$, the supersymmetric condition for a size modulus
is satisfied with a large value of $V$.

In both cases, the non-perturbative contribution to superpotential is
of the form $W_{\rm np}=ae^{-bX}$ (for the case (i), the leading order
in the large volume limit). The effective superpotential is therefore
given by $W=W_0+W_{\rm np}$. It is found that all the complex
structure moduli and the ten-dimensional dilaton are frozen by the
fluxes (the first term), which are essentially decoupled from 
low-energy effective theory and thus do not play any role in
supersymmetry breaking, while the second term generated by
non-perturbative stringy corrections can stabilize all remaining
K\"ahler moduli which arise in low-energy effective theory. As we
mentioned, $W_0$ is a constant with respect to the K\"ahler size
moduli. Moreover the parameter $b$ in $W_{\rm np}$ is real in either
case. Remarkably, the resulting moduli superpotential is exactly the
same form as we have derived in the previous subsection by looking for
phenomenologically preferable moduli potential.

In heterotic/M theory, there might also exist similar mechanism to
stabilize all heterotic moduli to the proposal of KKLT\@. First,
flux-induced superpotentials fix all shape moduli and in some cases
also the size modulus, which is not the case for type IIB
string. After including non-perturbative effects such as gaugino
condensation and world-sheet instanton effects, all remaining moduli
are shown to be stabilized~\cite{hetero}. See also other approaches,
for example, using non-K\"ahler background with gaugino condensation to
stabilize the dilaton modulus~\cite{nonKahler}. Anyway the resulting
superpotential generally takes the common form, $W=W_0+W_{\rm np}$, and
is consistent to our result (\ref{spot1}) of CP-conserving moduli
potential (now for the dilaton or radius modulus).

We finally comment on typical scales of dimensionful parameters as
well as possible field-theoretical origins of the moduli
superpotential. The exponential dependence of the moduli field is
realized in various ways. As explained above, for the dilaton and
related variables, that is induced by gaugino condensations or
non-perturbative effects in supersymmetric gauge theories. Therefore a
natural scale of the coefficient $a$ in (\ref{spot1}) might be on the
fundamental scale of the theory. On the other hand, the constant 
term $c$ is related to supersymmetry breaking and should be suppressed
relative to the fundamental scale in order to have low-energy
supersymmetry. The detailed analysis in the following sections will
show that a small $c$ is, in fact, relevant to various phenomenology of
the model. In the flux compactification, following the spirit of 
Refs.~\cite{landscape}, a small constant $c$ may be obtained by
cancellation among contributions from many fluxes. The suppression can
also be achieved by similar dynamics which generate the exponential
term, that is, an exponentially small $c$ is obtained by gaugino
condensation in additional gauge factor. Another simple way to have
a suppressed constant term comes from symmetry argument and is
described by a Wess-Zumino model without introducing any extra gauge
factors. Let us consider the following superpotential:
\begin{equation}
  W_c \,=\, \lambda\psi(\bar\phi\phi-M^2)+m\bar\phi\phi.
  \label{Wc}
\end{equation}
The last term in the right-hand side causes soft breaking 
of $R$ symmetry, and one expects a constant superpotential to be
generated. Integrating out the massive fields, we actually obtain a
constant superpotential $W_c=mM^2$. A suppressed value of $c$ is
natural in a sense that the first and second terms in (\ref{Wc}) 
are $R$ exact and the third softly breaks the $R$ invariance. It is
noted that the superpotential (\ref{Wc}) is the most general and
renormalizable one, provided that one imposes the 
parity [$P(\psi)=+$, $P(\phi,\bar\phi)=-$] in addition to 
the $R$ invariance [$R(\psi)=2$, $R(\phi,\bar\phi)=0$].

%%%%%%%%%%%%%%%%%%%%%%%%%%%%%%%%%%%%%%%%%%%%%%%%%%%%%%%%%%%%%%%%%%%%%%
\section{Moduli stabilization}
%%%%%%%%%%%%%%%%%%%%%%%%%%%%%%%%%%%%%%%%%%%%%%%%%%%%%%%%%%%%%%%%%%%%%%
\subsection{KKLT-type superpotential}
%%%%%%%%%%%%%%%%%%%%%%%%%%%%%%%%%%%%%%%%%%%%%%%%%%%%%%%%%%%%%%%%%%%%%%
\subsubsection{Potential analysis}

In the previous section, we have found that a moduli 
superfield $X$ is expected to have the following potentials for
realistic phenomenology:
\begin{equation}
  K \,=\, K(X+X^\dagger), \quad \quad W(X) \,=\, ae^{-bX}+c.  
\end{equation}
The superpotential form has been determined by the requirement of
suppressing gaugino mass phases. In this section III, we will fully
analyze the supergravity potential of moduli and investigate
its phenomenological implications in the vacuum. The supergravity
potential in the conformal frame now to be investigated is given by
\begin{equation}
  V \,=\, e^{K/3}K_{X\bar X}^{\,-1}
  \left[\,|a|^2\big(b^2-bK_X-\kappa\big)e^{-b(X+X^\dagger)}
  +|c|^2(bK_X-\kappa)-\kappa\big(ac^*e^{-bX}\!+{\rm h.c.}\big)\,\right],
\end{equation}
where we have defined the real 
function $\kappa\equiv 3K_{X\bar X}-K_XK_{\bar X}+bK_X$. Note that
only the last term depends on the imaginary part $X_I$ of the moduli
field. Minimizing the potential with respect to $X_I$, we have
\begin{equation}
  0 \,=\, \frac{\partial\,V}{\partial X_I} \,=\,
  b\kappa e^{K/3}K_{X\bar X}^{\,-1}\big(iac^*e^{-bX}+{\rm h.c.}\big),
\end{equation}
which determines the expectation value of $X_I$ as
\begin{equation}
  a^*ce^{ibX_I} \,=\, \epsilon|ac| \qquad (\epsilon=\pm1).  
\end{equation}
The sign of $\epsilon$ is chosen so that the extremum is a minimum,
that is, the second derivative with respect to $X_I$ should be
positive at the extremum. This results in the 
condition $\epsilon\kappa>0$. In the minimum of $X_I$, the moduli
potential becomes
\begin{equation}
  V(X_R) \,=\, e^{K/3}K_{X\bar X}^{\,-1}\Big[
  \,|a|^2\big(b^2-bK_X-\kappa\big)e^{-2bX_R} +|c|^2(bK_X-\kappa)
  -2|ac|\epsilon\kappa e^{-bX_R} \Big].
\end{equation}
The analysis of $V(X_R)$ is performed with specifying the moduli
K\"ahler potential. In this paper, we take a simple assumption
\begin{equation}
  f \,=\, -3(X+X^\dagger)^{n/3},
  \label{fX}
\end{equation}
often discussed in the literature. For example, in string effective
theory framework, $n$ is an integer ($n=1,2,3$). Having this K\"ahler
potential at hand, we find the minimum for a suppressed value
of $c$, which is related to the gravitino mass scale. It is important
that the minimization leads to a large value of $bX_R$ ($>0$), and
also a negative $\epsilon$ from the minimum 
condition ($\epsilon\kappa>0$). We thus obtain the stabilized value of
moduli which satisfies the following equation at the minimum
\begin{equation}
  e^{-bX_R} \,=\, \frac{n}{2}\left|\frac{c}{a}\right|\frac{1}{bX_R}
  \bigg[1-\frac{n+2}{2}\frac{1}{bX_R} 
  +{\cal O}\big((bX_R)^{-2}\big)\bigg].
\end{equation}
One can easily see that the second derivative with respect to $X_R$ is
certainly positive at this extremum. The detailed analysis of the
global structure of moduli potential will be performed in a later
section where the hidden-sector contribution and thermal effects in
the early Universe are included into the potential.

%%%%%%%%%%%%%%%%%%%%%%%%%%%%%%%%%%%%%%%%%%%%%%%%%%%%%%%%%%%%%%%%%%%%%%
\subsubsection{F terms and the vacuum energy}

In the obtained vacuum, supersymmetry is broken due to the presence of
the constant factor $c$. The non-vanishing $F$ terms are evaluated
from the general formulas (\ref{F_X}) and (\ref{F_phi}):
\begin{eqnarray}
  F_X &\,=\,& \frac{2c^*}{b(2X_R)^{n/3}}
  \left(1-\frac{2n-1}{2}\frac{1}{bX_R}+\cdots\right), \\
  F_\phi \,&\,=\,& \frac{c^*}{(2X_R)^{n/3}}
  \left(1-\frac{5n}{6}\frac{1}{bX_R}+\cdots\right). 
\end{eqnarray}
The supersymmetry-breaking scale is controlled by $c$. In the 
limit $c\to0$, the vacuum goes to infinity and supersymmetry is
restored. Thus the constant term $c$, which is allowed in a generic
solution (\ref{spot1}), is found to play important roles for the
moduli stabilization and also moduli supersymmetry breaking. In the
following discussion of moduli phenomenology, a key ingredient is the
ratio of two $F$-term contributions, which is now given by
\begin{equation}
  \frac{(F_X/X_R)}{F_\phi} \,=\, \frac{2}{bX_R}
  \left(1-\frac{n-3}{6}\frac{1}{bX_R}+\cdots\right),
\end{equation}
where the ellipsis denotes real factors 
of ${\cal O}((bX_R)^{-2})$. We find in the vacuum that the moduli
contribution to supersymmetry breaking is suppressed, by a large value
of $bX_R$, relative to the gravity contribution of $F_\phi$. The
anomaly mediated contribution to visible fields is also suppressed, by
one-loop factors, relative to $F_\phi$. It is interesting that the
supersymmetry breaking effects mediated by the moduli and conformal
anomaly are comparable in size in the heavy gravitino scenario. The
theory predicts neither pure anomaly mediation nor pure moduli
dominance and has characteristic mass spectrum, as we will study in
the next section.

The above potential analysis has not been concerned about the vacuum
energy (the cosmological constant). At the minimum given above, 
the $F$ term of the $X$ field is suppressed, and thus the vacuum
energy is negative. We therefore need to uplift the potential to make
its vacuum expectation value zero. In the KKLT model~\cite{KKLT}, an
anti-D3 brane is introduced to break supersymmetry explicitly to
generate a positive contribution to the vacuum energy. Here we
introduce hidden-sector fields with nonzero vacuum expectation values
of $F$ terms. One can incorporate in the supergravity $f$ function the
direct couplings between hidden-sector variables and the moduli field,
without conflicting with viable phenomenology in the visible
sector. The perturbative expansion about a hidden-sector 
field $Z$ gives a K\"ahler potential
\begin{equation}
  f \;=\; -3(X+X^\dagger)^{n/3}+H(X+X^\dagger)|Z|^2+{\cal O}(|Z|^4),  
\end{equation}
with a real function $H$, while preserving the moduli dependence of
K\"ahler terms in the form $X+X^\dagger$. The first term is the moduli
K\"ahler potential at the leading order (\ref{fX}). To avoid
introducing an additional CP phase, we assume that the vacuum
expectation value of $Z$ is negligibly small compared to the Planck
scale. Minimizing the supergravity potential with respect to the
moduli $X$, we find that the vacuum is shifted by turning on $F_Z$, in
which vacuum the $F$ terms are given by
\begin{eqnarray}
  F_X &\,\simeq\,& \frac{2c^*}{b(2X_R)^{n/3}}
  +\frac{2H_XX_R}{nbc} |F_Z|^2, \\
  F_\phi \,&\,\simeq\,& \frac{c^*}{(2X_R)^{n/3}}.
\end{eqnarray}
The requirement of vanishing cosmological constant can be fulfilled
with a non-vanishing $F$ term of hidden-sector 
field; $|F_Z|^2\simeq 3|c|^2/(2X_R)^{n/3}H$ at the
minimum. Substituting it for $F_X$, we obtain the $F$ terms in the
shifted vacuum, in particular, whose ratio becomes in the leading order
\begin{equation}
  \frac{(F_X/X_R)}{F_\phi} \,\simeq\, \frac{2}{bX_R}
  \bigg(1+\frac{3X_RH_X}{nH}\bigg).
\end{equation}
For example, $H\propto(X+X^\dagger)^l$ gives a correction
$\frac{(F_X/X_R)}{F_\phi}=\frac{2}{bX_R}\big(1+\frac{3l}{2n}\big)$.
Thus the moduli mediated contribution is still smaller 
than $F_\phi$ by large $bX_R$ suppression. Moreover the ratio of 
two $F$ terms remains real and does not disturb the phase alignment of
supersymmetry-breaking soft mass parameters.

%%%%%%%%%%%%%%%%%%%%%%%%%%%%%%%%%%%%%%%%%%%%%%%%%%%%%%%%%%%%%%%%%%%%%%
\subsection{Racetrack-type superpotential}

Similar analysis can be performed for another solution of CP
conservation, i.e., the racetrack type superpotential of moduli field
(\ref{spot2}):
\begin{equation}
  W(X) \,=\, a_1e^{-b_1X}+a_2e^{-b_2X}.
\end{equation}
We here take $|b_1|>|b_2|$ without loss of any generalities. If the
moduli is stabilized with interplay between these two terms, 
low-energy supersymmetry requires large [${\cal O}(10)$] values both
for $b_1X_R$ and $b_2X_R$ or only for $b_1X_R$. In the latter case,
the coupling $a_2$ must be hierarchically suppressed. That is
qualitatively very similar to the case of superpotential (\ref{spot1})
in the previous subsection with $c$ being replaced 
by $a_2e^{-b_2X}\sim a_2$. A straightforward calculation, in fact,
indicates that the stabilization and $F$-term equations are almost the
same also quantitatively. Therefore it is sufficient to consider the
case that both $b_1X_R$ and $b_2X_R$ become large. The minimization
with respect to $X_I$ then leads to the 
condition $a_1^*a_2e^{i(b_1-b_2)X_I}=-|a_1a_2|$. We assume the
moduli K\"ahler potential (\ref{fX}) now for the superpotential
(\ref{spot2}). In this case, the moduli is stabilized at the minimum
\begin{equation}
  e^{-(b_1-b_2)X_R} \,=\, \left|\frac{a_2}{a_1}\right|
  \left[\frac{b_2}{b_1}+\frac{n(b_1-b_2)}{2b_1}\frac{1}{b_1X_R}
  +{\cal O}\big((b_1X_R)^{-2}\big)\right],
\end{equation}
and the $F$ terms read from the general formulas as
\begin{eqnarray}
  F_X &\,=\,& \frac{2na_2^*e^{-b_2X^*}}{b_1b_2(2X_R)^{n/3+1}}
  \frac{b_1-b_2}{b_1}\left(1-\frac{n(b_1+2b_2)}{2b_2}\frac{1}{b_1X_R}
  +\cdots\right), \\
  F_\phi \,&\,=\,& \frac{a_2^*e^{-b_2X^*}}{(2X_R)^{n/3}}
  \frac{b_1-b_2}{b_1}\left(1-\frac{n}{2}\frac{1}{b_1X_R}+\cdots\right).
\end{eqnarray}
It may be interesting to see that the ratio of $F$ terms becomes
\begin{equation}
  \frac{(F_X/X_R)}{F_\phi} \,=\, \frac{n}{b_1b_2X_R^2}
  \left(1-\frac{n(b_1+b_2)}{2b_1b_2X_R}+\cdots\right).
\end{equation}
Therefore the moduli become much heavier than the gravitino in
contrast to the case with the superpotential (\ref{spot1}), and the
moduli mediated contribution to supersymmetry breaking is highly
suppressed. This means that the superparticle mass spectrum is described
dominantly in terms of the anomaly mediation. As is well known, that
makes some scalar leptons tachyonic as long as low-energy theory
is the minimal supersymmetric standard model without some additional
sources of supersymmetry breaking. On these grounds, we will focus in
this paper on the moduli superpotential of the form (\ref{spot1}),
with which potential the moduli physics gives rise to rich and
significant effects on low-energy phenomenology.

%%%%%%%%%%%%%%%%%%%%%%%%%%%%%%%%%%%%%%%%%%%%%%%%%%%%%%%%%%%%%%%%%%%%%%
\section{Mass spectroscopy}
%%%%%%%%%%%%%%%%%%%%%%%%%%%%%%%%%%%%%%%%%%%%%%%%%%%%%%%%%%%%%%%%%%%%%%
\subsection{Gravitino and moduli masses}

In the previous section, we considered the following K\"ahler and
superpotential of moduli and hidden-sector fields:
\begin{equation}
  f \;=\; -3(X+X^\dagger)^{n/3} + (X+X^\dagger)^l |Z|^2, \qquad
  W(X) \,=\, a e^{-bX} + c, 
  \label{fW}
\end{equation}
where $n = 1,2,3$. With these forms at hand, we simultaneously
obtained the stabilized moduli, broken supersymmetry, and the
vanishing cosmological constant in the vacuum of supergravity
potential. There exist two types of supersymmetry-breaking 
contributions, $F_X/X_R$ and $F_\phi$, in this model. In particular,
the superpotential puts these $F$ terms to be not independent at the
vacuum and approximately satisfy the relation
\begin{equation}
  \frac{F_X}{X_R} \,=\, \frac{2}{bX_R} 
  \bigg(1+\frac{3l}{2n}\bigg) F_\phi.
  \label{FXFphi}
\end{equation}
It is important to notice that, in the vacuum, the leading terms 
in $F_X$ are canceled out in the large $bX_R$ limit, and the moduli 
contribution $F_X/X_R$ is suppressed by $bX_R$, compared  
to $F_\phi$. Then the moduli and hidden-sector $F$ terms are found to
be small at the vacuum, and the gravitino mass is expressed as
\begin{equation}
  m_{3/2} \,=\, |F_\phi|.
\end{equation}

The moduli field is stabilized by non-perturbative effect plus the
constant superpotential, latter of which is a trigger of supersymmetry
breaking.  Therefore a natural expectation might be that its mass is
around the symmetry-breaking scale. We however obtain the moduli mass
which reads from the Lagrangian (\ref{L})
\begin{equation}
  m_{X_R} \,=\, m_{X_I} \,=\, 2\sqrt{2}\, bX_R m_{3/2},
\end{equation}
after making moduli-dependent Weyl rescaling to go from the
superconformal frame to the Einstein frame.

A rough estimation of the mass spectrum of the theory becomes as
follows: we have found in the above that the moduli is one or two
orders of magnitude heavier than the gravitino by a large factor 
of $bX_R\sim{\cal O}(10)$, and is one of the heaviest fields in the
low-energy spectrum, together with its axionic scalar 
partner $X_I$. Such heavy moduli can be a solution to the moduli
problem and also have interesting cosmological implications discussed
in later sections (see also~\cite{KYY1,KYY2}). The next-to-heaviest
particle is the gravitino whose mass $F_\phi$ is no less 
than $O(10)$~TeV, which is parameterized in our model by a constant
factor $c$. This mass scale is suitable to unravel the cosmological
gravitino problem. Although the moduli decouples at a high scale, its
threshold is given by $F_\phi$. Hence the corrections from the moduli
can alter the pure anomaly mediated spectrum of superparticles. In
fact, the $F_X$ contribution becomes comparable to that from the
super-Weyl anomaly. As we will explicitly see below,
superparticles in the visible sector, squarks, sleptons, and gauginos,
are yet lighter than the gravitino by one-loop factors 
or $bX_R$ suppressions.

%%%%%%%%%%%%%%%%%%%%%%%%%%%%%%%%%%%%%%%%%%%%%%%%%%%%%%%%%%%%%%%%%%%%%%
\subsection{Gaugino masses}

As have been already derived in (\ref{gauginomass}), a gaugino mass
receives the moduli and anomaly mediated contributions:
\begin{equation}
  {M_\lambda}_i  \,=\, 
  \frac{F_X}{2X_R}+\frac{{\beta_g}_i}{g_i}F_\phi \,=\,
  \left(\frac{1+\frac{3l}{2n}}{bX_R}
    +\frac{\beta_i g_i^2}{16\pi^2}\right)F_\phi,
  \label{gaumass}
\end{equation}
where we have taken into account the one-loop order expression for the
anomaly mediation and also used the relation between the $F$ terms in
the vacuum. The constants $\beta_i$ denote the beta function 
coefficients. We have dropped direct hidden-sector contribution, which
may be easily justified by assuming that hidden-sector fields
participating in supersymmetry breaking are charged under some gauge
or global symmetry. The gaugino mass spectrum has several
characteristic features, predicted by the existence of two sources of
supersymmetry breaking, which we will see by applying it to the
minimal supersymmetric standard model.

For later convenience, we define the quantity $R_{bX}$ as
\begin{equation}
   R_{bX} \,\equiv\, \frac{F_{\phi}}{F_X/2 X_{R}} 
   \,=\, \frac{b X_R}{1+\frac{3l}{2n}}.
\end{equation}
Then Eq.~(\ref{gaumass}) reads
\begin{equation}
  {M_\lambda}_i \,=\, 
  \left(1+\frac{\beta_i g_i^2}{16\pi^2}\cdot R_{bX}\right)
  \frac{F_X}{2X_R} \,=\, \left(\frac{1}{R_{bX}}
    +\frac{\beta_i g_i^2}{16\pi^2}\right)F_\phi.
\end{equation}

%%%%%%%%%%%%%%%%%%%%%%%%%%%%%%%%%%%%%%%%%%%%%%%%%%%%%%%%%%%%%%%%%%%%%%
\subsection{Scalar masses}
%%%%%%%%%%%%%%%%%%%%%%%%%%%%%%%%%%%%%%%%%%%%%%%%%%%%%%%%%%%%%%%%%%%%%%
\subsubsection{Practical K\"ahler potential for visible scalars}

Visible sector scalars such as squarks and sleptons generally receive
non-holomorphic supersymmetry-breaking masses from their couplings to
supersymmetry-breaking fields. The visible scalar mass spectrum is
rather sensitive to detailed form, in particular, of their K\"ahler
potential. Therefore feasible field dependence of K\"ahler potential
is highly restricted in constructing phenomenologically and
theoretically viable models. In this subsection, paying particular
attention to sfermion masses, we first clarify what form of 
visible-sector K\"ahler potential should be in the heavy gravitino
scenario.

For chiral supermultiplets, the most general supergravity Lagrangian
is given by two ingredients: K\"ahler potential $K$ and 
superpotential $W$ [see Eq.~(\ref{L})]. In what follows, we examine 
the supergravity $f$ function ($f=-3e^{-K/3}$), instead of $K$, as it 
may be easier to analyze scalar potentials in the conformal frame of
supergravity. To see physical soft masses of visible scalars $Q$, we
first integrate out the auxiliary components $F_Q$ via their equations
of motion
\begin{equation}
  F_\phi^* f_Q+W_Q +\sum_I F_I^\dagger f_{Q\bar I} \,=\, 0,
\end{equation}
where the lower indices of $f$ and $W$ denote the field
derivatives. The index $I$ runs over all chiral multiplet scalars in
the theory. After the integration, the resultant scalar potential is
given by
\begin{eqnarray}
  -V &\,=\,& \bigg(f-\frac{f_Qf_{\bar Q}}{f_{Q\bar Q}}\bigg) 
  F_\phi^*F_\phi +\sum_{I\neq Q}\bigg(f_I-
  \frac{f_Qf_{I\bar Q}}{f_{Q\bar Q}}\bigg) F_\phi^*F_I +{\rm h.c.} 
  +\!\sum_{I,\,J\neq Q} \!\bigg(f_{\bar IJ}-
  \frac{f_{Q\bar I}f_{J\bar Q}}{f_{Q\bar Q}}\bigg) 
  F_I^\dagger F_J  \nonumber \\
  && \qquad +\bigg(3WF_\phi +\sum_{I\neq Q}W_IF_I -f_{Q\bar Q}^{-1}
  \bigg[\frac{1}{2}|W_Q|^2 +W_Q\Big(F_\phi f_{\bar Q} +
  \sum_{I\neq Q}F_If_{I\bar Q}\Big)\bigg] +{\rm h.c.}\bigg).
  \label{VQ}
\end{eqnarray}
It turns out that visible scalar masses do not appear from the second
line of the potential (\ref{VQ}) unless the superpotential $W$
contains mass terms of visible scalar multiplets.

Let us examine supersymmetry-breaking masses of $Q$ and resultant
forms of visible scalar K\"ahler potential. We first require that, in 
the heavy gravitino scenario, the tree-level gravity~($F_\phi$)
contribution should be suppressed. If this is not the case, large
flavor violation is generally induced via superpartners of quarks and
leptons without invoking some additional mechanism to guarantee flavor
conservation. Further, as for the Higgs fields, large $F_\phi$ effects
prevent their required expectation values from being generated. Thus
we have the following phenomenological requirement that the first two
terms in (\ref{VQ}) do not contain non-holomorphic mass 
of $Q\,$:\footnote{Exactly speaking, the most general conditions
become $f-\frac{f_Qf_{\bar Q}}{f_{Q\bar Q}}=
A(Q)+\bar A(\bar Q)+(Q\textrm{-free terms})$, etc.\ where $A$ is an
arbitrary function.}
\begin{equation}
  \qquad f-\frac{f_Qf_{\bar Q}}{f_{Q\bar Q}} \,=\, (Q\textrm{-free}), 
  \qquad f_I-\frac{f_Qf_{I\bar Q}}{f_{Q\bar Q}} \,=\, 
  (Q\textrm{-free}), \quad (I\neq Q,\; F_I\neq0).
\end{equation}
We find the generic solution
\begin{equation}
  f \,=\, L(I,I^\dagger)M(Q,I)\bar M(Q^\dagger\!,I^\dagger) 
  +N(I,I^\dagger),
  \label{sol1}
\end{equation}
with arbitrary functions $L$, $M$ and $N$. ($L$ and $N$ should be real
functions.) \ It is clear to see with this solution that 
tree-level $F_\phi$ effects disappear since the compensator dependence
can be removed from the visible-sector K\"ahler potential by 
holomorphic field redefinition $\Phi M\to M$. In the 
solution (\ref{sol1}), the symbol $I$ means supersymmetry-breaking
superfields ($F_I\neq 0$) such as hidden sector variables $Z$ and
modulus fields $X$. If there are some hidden-sector multiplets with 
non-vanishing $F$ components, their direct couplings to the visible
sector should also be suppressed to avoid disastrous flavor-violating
phenomena. Similar to the above conditions, it is sufficient to
require that the coefficients of $F_Z^\dagger F_I$ in the 
potential (\ref{VQ}) do not depend on visible-sector scalars:
\begin{equation}
  f_{Z\bar I}-\frac{f_{Z\bar Q}f_{Q\bar I}}{f_{Q\bar Q}}
  \,=\, (Q\textrm{-free}),
\end{equation}
where $I$ denotes all the chiral multiplets whose $F$ components have
non-vanishing expectation values. This condition can be solved,
combined with Eq.~(\ref{sol1}), and we obtain
\begin{equation}
  f \,=\, L(X,X^\dagger)M(Q,Z)\bar M(Q^\dagger\!,Z^\dagger) 
  +N(X,X^\dagger\!,Z,Z^\dagger),
  \label{sol2} 
\end{equation}
with arbitrary functions $L$, $M$ and $N$. The $X$ dependence has
dropped out from the function $M$ because the K\"ahler potential
depends on moduli fields in the form of $X+X^\dagger$. With the
K\"ahler potential (\ref{sol2}), visible-sector scalars $Q$ receive
tree-level non-holomorphic soft masses only from the moduli 
fields ($m_Q^2\propto|F_X|^2$). The function $N$ includes the moduli
and hidden-field K\"ahler potential, e.g.\ Eq.~(\ref{fW}) we discussed
before. It might be interesting to notice that the sequestering of
visible sector ($Q$) from the hidden one ($Z$) is not exact 
in (\ref{sol2}). But the hidden-sector fields $Z$ do not induce any
tree-level (and potentially dangerous) masses of visible scalars.

Finally, what is the form of K\"ahler potential such that
visible-sector scalars do not receive any non-holomorphic soft masses
even from moduli fields? That results in a requirement that the 
coefficient of $|F_X|^2$ in the potential (\ref{VQ}) does not depend
on $Q$:
\begin{equation}
  f_{X\bar X}-\frac{f_{X\bar Q}f_{Q\bar X}}{f_{Q\bar Q}} 
  \,=\, (Q\textrm{-free}).
\end{equation}
We find that this requirement determines the functional form 
of $L(X,X^\dagger)$ in (\ref{sol2}):
\begin{equation}
  f \,=\, e^{k(X+X^\dagger)}M(Q,Z)\bar M(Q^\dagger\!,Z^\dagger) 
  +N(X,X^\dagger\!,Z,Z^\dagger),
  \label{most}
\end{equation}
with arbitrary functions $M$ and $N$, and $k$ being a real
constant. With this most limited K\"ahler form at hand, visible-sector
scalars obtain no non-holomorphic supersymmetry-breaking masses at
classical level. The supergravity $f$ function (\ref{most}) has a
special and interesting property. A limited case of $k=0$ is the
well-known `no-scale' type K\"ahler potential. Namely, the visible
sector and moduli dynamics are separated in the K\"ahler potential. 
For $k\neq 0$, visible scalar masses also vanish in spite of the fact 
that the K\"ahler potential is no longer of a sequestered type. It may
be worth mentioning that there is a realistic model in which this type
of K\"ahler potential is dynamically realized. That is the
supersymmetric warped extra dimensions~\cite{SUSYRS} where the
constant $k$ is proportional to the anti de-Sitter 
curvature, and $X$ is the radius modulus which determines the size of
compact space. In this case, vanishing scalar masses may be understood
as a result of scale invariance in four-dimensional conformal field
theory dual to the original five-dimensional one. Whether other
dynamics, especially some four-dimensional models, leads to the
K\"ahler potential (\ref{most}) might deserve to be investigated.

%%%%%%%%%%%%%%%%%%%%%%%%%%%%%%%%%%%%%%%%%%%%%%%%%%%%%%%%%%%%%%%%%%%%%%
\subsubsection{Supersymmetry-breaking scalar soft terms}

Having established the possible K\"ahler potential for visible
scalars, we hereafter consider the following tree-level 
supergravity $f$ function:
\begin{equation}
  f \;=\; -3(X+X^\dagger)^{n/3} + (X+X^\dagger)^l |Z|^2
  +\sum_i (X+X^\dagger)^{m_i}|Q_i|^2,
  \label{fXZQ}
\end{equation}
where $X$ is the modulus, $Z$ the hidden-sector variable, and $Q_i$'s
the visible-sector ones such as squarks and sleptons. We have assumed
for simplicity the sequestering of the visible and hidden
sectors~\cite{IKYY}, but will come back to this point later on. It is
found from (\ref{sol2}) that the visible scalars receive tree-level
non-holomorphic soft masses only from the moduli 
contribution $F_X$. At quantum level, the anomaly 
mediation ($F_\phi$) induces additional scalar mass terms. As in the
gaugino mass, the $\Phi$-dependent pieces come out through the
renormalization in the combination of $\mu/\Lambda|\Phi|$. In addition
to the potential terms (\ref{VQ}), several $\phi$-derivative K\"ahler 
terms generally appear, which are possible sources of visible scalar
soft masses:
\begin{equation}
  \bigg[ \bigg(f_{I\bar\phi}
  -\frac{f_{Q\bar\phi}f_{I\bar Q}}{f_{Q\bar Q}}\bigg)F_\phi^*F_I
  +{\rm h.c.}\bigg] 
  +\bigg[\bigg(f_\phi-\frac{f_Qf_{\phi\bar Q}}{f_{Q\bar Q}}\bigg)
  +{\rm h.c.} +\bigg(f_{\phi\bar\phi}
  -\frac{f_{Q\bar\phi}f_{\phi\bar Q}}{f_{Q\bar Q}}\bigg)\bigg]F_\phi^*F_\phi.
\end{equation}
The latter part gives rise to the usual anomaly mediated
contribution and the former one includes the moduli-anomaly mixing
pieces for $I=X$. Here the hidden-sector contribution is suppressed
due to small expectation values of $Z$'s. 
Neglecting higher-order loop corrections, we obtain the total soft
masses of visible scalars at the leading order:
\begin{eqnarray}
  {m_Q^2}_i &\,=\,& m_i\left|\frac{F_X}{2X_R}\right|^2
  -\bigg(\frac{1}{2}\frac{\partial\gamma_i}{\partial X_R}
  F_\phi^*F_X +{\rm h.c.}\bigg)
  +\frac{1}{2}\frac{\partial\gamma_i}{\partial\ln\mu}\,|F_\phi|^2 
  \nonumber\\
  &\,=\,& \bigg(\frac{m_i}{R_{bX}^{\,2}} 
  -\frac{2}{R_{bX}}\frac{\partial \gamma_i}{\partial\ln X_R}
  +\frac{1}{2}\frac{\partial\gamma_i}{\partial\ln\mu}\bigg)|F_\phi|^2,
  \label{mQ}  
\end{eqnarray}
where $\gamma_i$ is the anomalous dimension of corresponding
scalar, and its derivative with respect to $X_R$ is given by
\begin{equation}
  \frac{\partial \gamma_i}{\partial\ln X_R}
  \,=\, -\sum_{j,k}(m_i+m_j+m_k)r|y_{ijk}|^2+sg^2
  \frac{\partial\ln(1/g^2)}{\partial\ln X_R},
\end{equation}
for the tree-level K\"ahler form (\ref{fXZQ}) 
and $\gamma_i=r|y_{ijk}|^2-sg^2$ where $r$ denotes the multiplicity
of Yukawa coupling in the one-loop anomalous dimension and $s$ is
determined by the quadratic Casimir of the $i$ field for a gauge group
with the gauge coupling $g\,$ ($r,s\geq0$). The indices $j,\,k$ in the
summation denote the superfields which couple to the $i$ field with
the Yukawa coupling $y_{ijk}$. In the above mass formula,
flavor-mixing couplings have not been included in $\gamma$'s, but can
easily be incorporated in a straightforward manner. It is noticed that
in the vacuum of the theory, two $F$ terms are correlated to each
other and all the contributions in (\ref{mQ}) are comparable in size
for a large value of $R_{bX}$.

Visible-sector scalars also have supersymmetry-breaking holomorphic
terms, including tri-linear and bi-linear ones in scalar 
fields (usually called $A$ and $B$ terms, respectively). These are
generated in the presence of superpotential $W(Q)$ of visible-sector
chiral multiplets. The general tree-level expression of holomorphic
terms is calculated from the scalar potential (\ref{VQ}):
\begin{equation}
  \qquad {\cal L}_h \,=\, -\frac{F_X}{2X_R}\sum_im_iQ_iW_{Q_i}(Q)
  +F_\phi\Big[3W(Q)-\sum_iQ_iW_{Q_i}(Q)\Big] \,+{\rm h.c.}.
\end{equation}
The subscripts of $W$ denote the field derivatives and the sum 
of $i$ are taken for all visible scalars in the theory. As an example,
let us consider Yukawa and supersymmetric mass 
terms; $W(Q)=y_{ijk}Q_iQ_jQ_k+\mu_{ij}Q_iQ_j$. The induced tri-linear
and bi-linear scalar couplings are read from the general 
expression ${\cal L}_h$, and the leading order expressions are given by
\begin{eqnarray}
  \qquad A_{ijk} &\,=\,& \frac{F_X}{2X_R}(m_i+m_j+m_k)
  -F_\phi(\gamma_i+\gamma_j+\gamma_k), \label{A} \\
  B_{ij} \,&\,=\,& \frac{F_X}{2X_R}(m_i+m_j)
  -F_\phi(1+\gamma_i+\gamma_j).
\end{eqnarray}
We have defined as usual the soft-breaking 
parameters $A$'s and $B$'s as the ratios of couplings between
supersymmetric and corresponding supersymmetry-breaking terms. It is
important to notice that supersymmetry-breaking parameters are
described by $F_X$ and $F_\phi$ with some real coefficients. Though
these parameters are generally complex, all their phases are
dynamically aligned to gaugino mass phases in our model and can be
rotated away with a suitable $R$ rotation. The above $B$-term formula,
when applied to the minimal supersymmetric standard models and beyond,
causes a too large value of $B$ parameter to trigger correct
electroweak symmetry breaking, if the anomaly 
mediation ($F_\phi$ term) is dominant. Though there have been 
several solutions to this problem~\cite{Bmu}, they are highly
model-dependent and generally predict different values 
of $B$ according to how to develop $\mu$ parameters. We will later 
treat $\mu$ and $B$ of the Higgs doublets as dependent parameters
fixed by the conditions for the proper electroweak symmetry breaking.

We here comment on several representative values of $m_i$ and $n$ in
the K\"ahler potential. The most popular examples of moduli fields are
the dilaton and overall modulus in string-inspired four-dimensional
supergravity. The former determines the gauge coupling constant and
the latter fixes the size of six-dimensional compact space. The
K\"ahler terms of dilaton and overall modulus are known to 
have $n=1$ and $n=3$, respectively. The soft terms from these specific 
moduli were analyzed in Ref.~\cite{softterm} for heterotic string
models. In the limit of dilaton dominated supersymmetry breaking, all
field-dependences are dropped out and the spectrum becomes
universal. That is encoded to the case 
of $n=1$ and ${}^\forall m_i=1/3$. Another known limit is the moduli
dominated supersymmetry breaking 
where $n=3$ and $m_i=0$ ($m_i\leq-1$) for the large volume Calabi-Yau
models or for untwisted (twisted) sector fields in orbifold
compactification. A moduli dominated supersymmetry breaking 
with $n=3$ which fits with our consideration may be obtained in 
type IIB models with matter localized on D3/D7 branes. In fact, the
gauge fields on a D7 brane have a gauge kinetic function of the 
form $S(X)=X$, with $X$ being the overall size moduli, and the chiral
multiplets on D3 and D7 correspond to the 
cases $m_i=0$ and $m_i=1$, respectively~\cite{fluxinduce}.

In the numerical analysis below, we will take $l=m_i$, for
simplicity. Such a hidden-sector effect might be model-dependent but
is not expected to change qualitative results, even if included.

%%%%%%%%%%%%%%%%%%%%%%%%%%%%%%%%%%%%%%%%%%%%%%%%%%%%%%%%%%%%%%%%%%%%%%
\subsection{Application to the minimal supersymmetric standard model}

As the simplest example, we apply the general 
formulae (\ref{gaumass}), (\ref{mQ}), and (\ref{A}) to the minimal
supersymmetric standard model and examine the sparticle mass spectrum
predicted by the coexistence of moduli and anomaly mediated
supersymmetry breaking. At first, although there are multiple sources
of supersymmetry breaking, their complex phases can all be rotated
away by $R$ rotation. This is because of the dynamical alignment of CP
phases in the vacuum, as we have shown before. Thus our model is
survived by stringent phenomenological constraints from CP violation,
which come from, e.g.\ the electric dipole moments of charged leptons.

%%%%%%%%%%%%%%%%%%%%%%%%%%%%%%%%%%%%%%%%%%%%%%%%%%%%%%%%%%%%%%%%%%%%%%
\subsubsection{Gaugino masses}

\begin{figure}[htbp]
\begin{center}
\includegraphics[scale=1]{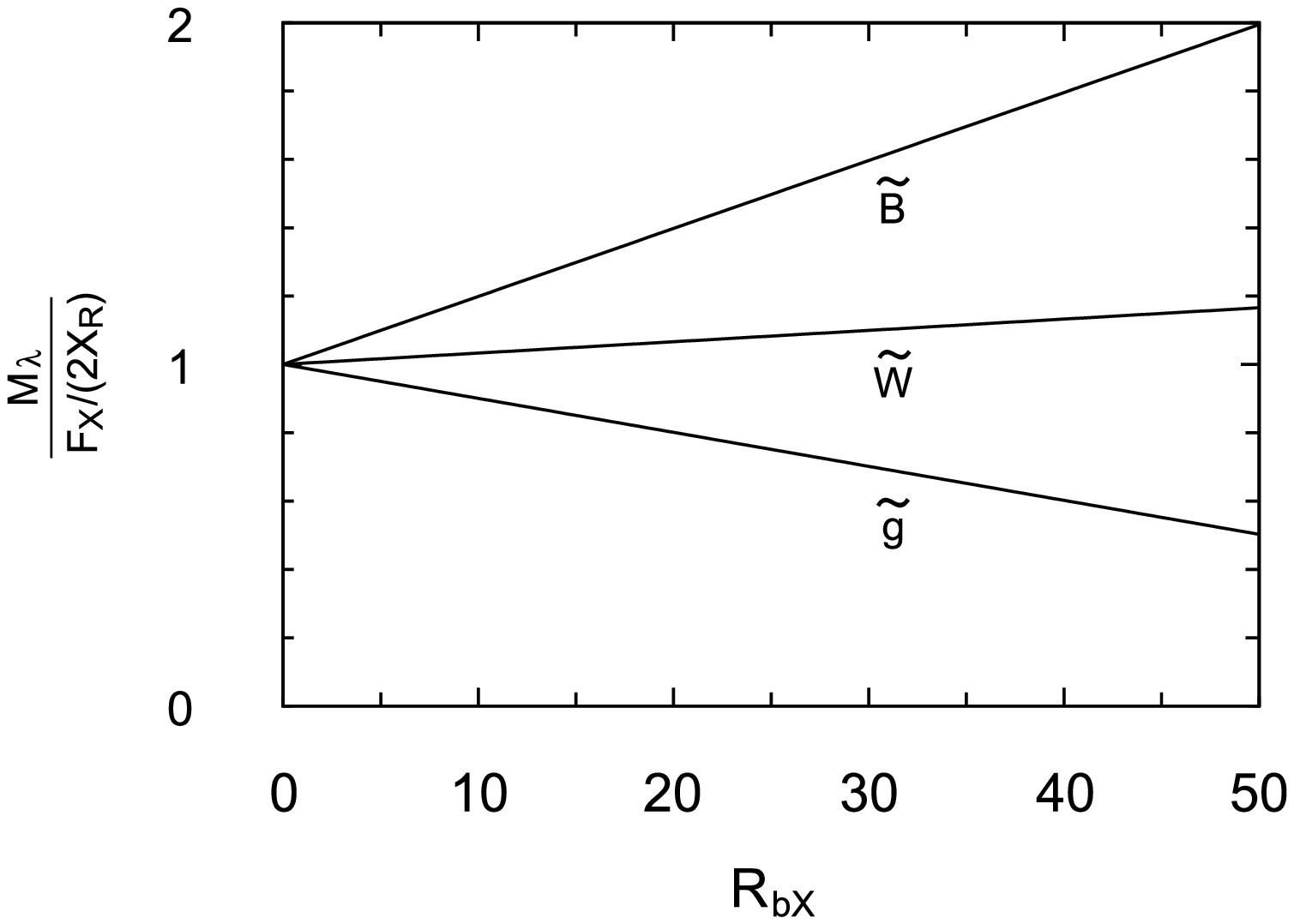}
\end{center}
\caption{Gaugino masses at the unification scale for the standard
model gauginos: gluino $\tilde g$, wino $\tilde W$, and 
bino $\tilde B$. Here we fix the moduli contribution $F_X/X_R$. The
region for small values of $bX_R$ is described by the extrapolation,
and the gaugino masses become universal at $bX_R=0$ where the moduli
contribution dominates the spectrum.}
\label{gaugino_GUT}
\end{figure}

The gaugino mass formula (\ref{gaumass}) contains the one-loop
anomaly-mediated contribution which is proportional to gauge beta
function and is given, for the minimal supersymmetric standard model,
by $\beta_i=(-3,1,33/5)$ for $SU(3)$, $SU(2)$, and $U(1)$,
respectively. Fig.~\ref{gaugino_GUT} shows typical gaugino masses at a
high-energy scale where the gauge couplings unify.  In the figure, we
fix the moduli contribution $F_X/X_R$. Therefore the slope of each
line is proportional to the gauge beta function. It is found that the
moduli mediated contribution, which is assumed to give unified gaugino
masses, dominates at the vanishing $R_{bX}$, whereas a larger value 
of $R_{bX}$ enhances the anomaly mediated effects and leads to larger
splitting of initial gaugino masses. We want to emphasize that this
result of non-unification spectrum is a rather generic prediction of
the scheme and is independent of the specific form of the moduli
K\"ahler potential. It is controlled only by the ratio $R_{bX}$.  Thus
the gaugino masses generally take the notable spectrum which is
different from the usual unification assumption. It is noted that if
there exists some unified gauge theory in a high-energy regime, the
standard model gaugino masses should be unified above the unification
scale. In this case, the gaugino mass splitting is caused by threshold
corrections at the symmetry-breaking scale of unified theory. This
type of scenario is compatible with our observation of non-universal
gaugino masses in the large $R_{bX}$ region.

\begin{figure}[htbp]
\begin{center}
\includegraphics[scale=1]{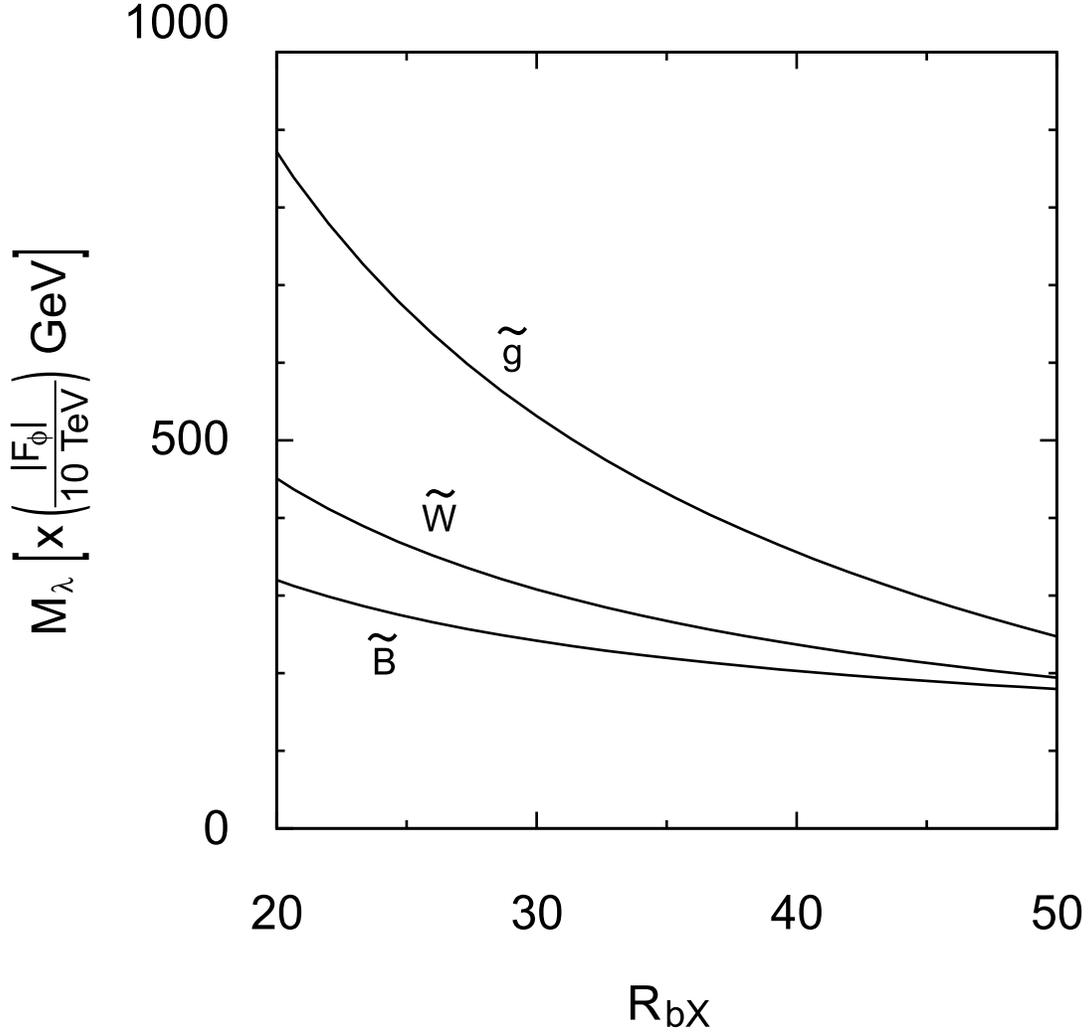}
\end{center}
\caption{Gaugino masses at the electroweak scale for the standard
model gauginos: gluino $\tilde g$, wino $\tilde W$, and 
bino $\tilde B$. Here we fix the anomaly contribution $F_\phi$ and
assume that the gaugino masses are mediated at the unification 
scale ($=2\times10^{16}$~GeV). For a larger value of $R_{bX}$, the
gaugino mass hierarchy decreases.}
\label{gaugino_EW}
\end{figure}

One of the most important consequences of the spectrum is thus the
mass degeneracy of gauginos at the electroweak scale. As seen 
in Fig.~\ref{gaugino_GUT}, for larger $R_{bX}$, the spectrum
significantly deviates from the universality at the cutoff 
scale: the $U(1)$ gaugino (bino) becomes heavier and 
the $SU(3)$ gaugino (gluino) becomes lighter, while 
the $SU(2)$ gaugino (wino) mass is not so affected. As a result, the 
three gaugino masses have a tendency to degenerate at the electroweak
scale after the renormalization-group effects are taken into
account. In Fig~\ref{gaugino_EW}, we describe the result of gaugino
masses at the electroweak scale. In the graph, we fix the 
compensator $F$ term ($F_\phi$). The initial gaugino masses are set
at the unification scale and are evaluated by using the one-loop
renormalization-group equations in the minimal supersymmetric standard
model. Then the lightest gaugino is found to be the bino for
reasonable range of parameter space, but the mass difference among the
three gauginos are rather reduced. The near degeneracy of low-energy
gaugino masses is an important prediction of our model, quite
different from the usual unification assumption at high-energy region,
e.g.\ as in supergravity models with hidden-sector supersymmetry
breaking.

%%%%%%%%%%%%%%%%%%%%%%%%%%%%%%%%%%%%%%%%%%%%%%%%%%%%%%%%%%%%%%%%%%%%%%
\subsubsection{General aspects of mass spectrum}

The mass spectrum of the matter and Higgs sectors can be read from
the general formulas (\ref{mQ}) and (\ref{A}), which come from the
K\"ahler potential (\ref{fXZQ}). Having at hand the similar order of
low-energy gaugino masses in the large $bX_R$ region, we generally
find several interesting aspects of sparticle spectrum:
\begin{itemize}
\item The colored superparticles (squarks) have similar magnitudes
of soft mass terms to those of non-colored particles (sleptons). This
is because the renormalization-group effects down to the electroweak
scale do not so much enhance the mass ratios of colored to non-colored
sfermions due to a suppressed initial value of gluino mass in
high-energy regime. That should be compared to the case of universal
hypothesis of gaugino masses, in which case, squarks generally become
much heavier than sleptons due to the strong gluino effect.
\item The mass scales of wino and bino are relatively large. This
behavior follows from the experimental lower bound on the lightest
Higgs boson mass $M_h$. The minimal supersymmetric standard model
gives a theoretical upper bound on $M_h\leq M_Z|\cos2\beta|$ at tree
level, where $M_Z$ is the $Z$ boson mass and $\tan\beta$ the ratio of
the vacuum expectation values of two Higgs doublets. However, this 
bound naively contradicts with the current lower limit from 
the LEP-II experiment of $M_h>114.4$~GeV~\cite{PDG}. The discrepancy
can be solved by large radiative corrections from the top
sector~\cite{HiggsRC}. The mechanism requires a relatively heavy mass
for the scalar partner of top quark, compared to the electroweak
scale. In a usual case, a heavy scalar top may be realized by the
strong gluino effects in renormalization-group evolution, while other
non-colored sparticles remains light around the electroweak
scale. However, in our model, colored and non-colored superparticles
have similar sizes of masses. Therefore a heavy scalar top, needed to
satisfy the Higgs mass bound, generally implies relatively heavy wino
and bino.
\item The fermionic partners of Higgs bosons (higgsinos) can be
relatively light. Their masses are controlled by the supersymmetric
Higgs mass parameter $\mu$. As mentioned before, we do not specify
dynamical origins of $\mu$ and corresponding $B$ parameter, and
instead fix them by the conditions for the electroweak symmetry
breaking. The behavior of the $\mu$ parameter can be understood by
considering the tree-level conditions. A positive $|\mu|^2$ is
realized by drawing down the up-type Higgs soft mass during the
renormalization-group running. Since the up-type Higgs is strongly
coupled to the scalar top, the gluino mass which controls the mass
scale of colored superparticles has a significant effect on the
low-energy $\mu$ parameter~\cite{KY}. We can see this behavior by
observing the following fitting formula:
\begin{eqnarray}
  |\mu|^2 &\,=\,& -0.006\,|{M_\lambda}_1|^2 -0.212\,|{M_\lambda}_2|^2 
  +1.54\,|{M_\lambda}_3|^2 \nonumber \\ 
  && \; +0.006\,|{M_\lambda}_1{M_\lambda}_2| 
  +0.139\,|{M_\lambda}_2{M_\lambda}_3| 
  +0.017\,|{M_\lambda}_3{M_\lambda}_1| \nonumber \\ 
  && \; -0.001\,|A_b|^2 +0.111\,|A_t|^2 
  +0.074\,|{M_\lambda}_2A_t|
  -0.007\,|{M_\lambda}_3A_b|
  +0.270\,|{M_\lambda}_3A_t| \nonumber \\ 
  && \; +0.339\,m_{\tilde q}^2 -0.001\,m_{\tilde b}^2
  +0.340\,m_{\tilde t}^2
  +0.009\,m_{H_d}^2 -0.670\,m_{H_u}^2 \nonumber \\ 
  && \; +0.026\, S_{\rm RGE} -M_Z^2/2.
  \label{mu-fitting}
\end{eqnarray}
Here the supersymmetry-breaking parameters in the right-hand side
are given at the unification scale of gauge 
couplings, and $|\mu|^2$ evaluated at 1~TeV\@. The formula is derived
by solving the renormalization group equations at the one-loop level
from the unification scale to 1~TeV, and using the tree-level Higgs
potential in the general minimal supersymmetric standard model. In the
estimation, we assume $\tan\beta=10$ and the top 
mass $M_t = 178$~GeV~\cite{topmass}. We notice 
that $S_{\rm RGE}$ arises due to the $U(1)$ hypercharge couplings with
the form, $S_{\rm RGE}={\rm Tr}(Y m^2)$, where the trace is
taken over all charged scalars in the model weighted by the
hypercharge $Y$. In the formula we find that the gluino mass effect is
dominant, while scalar mass contributions almost cancel each other,
which results in a tiny dependence in the estimation of $\mu$. In the
usual models with the universal gaugino mass hypothesis, the bino mass
is drawn down during the renormalization group running, so 
the $\mu$ parameter tends to be much larger than the bino 
mass (expect for the case of focus point~\cite{focus}). On the other
hand, in the present model, since the gluino mass is suppressed
compared to bino, the hierarchy becomes reduced. Actually $\mu$
becomes around the masses of bino and wino. Consequently, the lightest
neutralino contains a significant (or even dominant) component of
higgsino, which can be the lightest supersymmetric particle with
unbroken $R$ parity and become a candidate for cold dark matter in the
Universe. In the following analysis, we consider the Higgs potential
at the one-loop order. The numerical estimation including one-loop
corrections shows an agreement with the above fitting formula 
within $10$--$20$~\% accuracy.
\end{itemize}

%%%%%%%%%%%%%%%%%%%%%%%%%%%%%%%%%%%%%%%%%%%%%%%%%%%%%%%%%%%%%%%%%%%%%%
\subsubsection{Case study}

\begin{figure}[htbp]
\begin{center}
\includegraphics[scale=0.95]{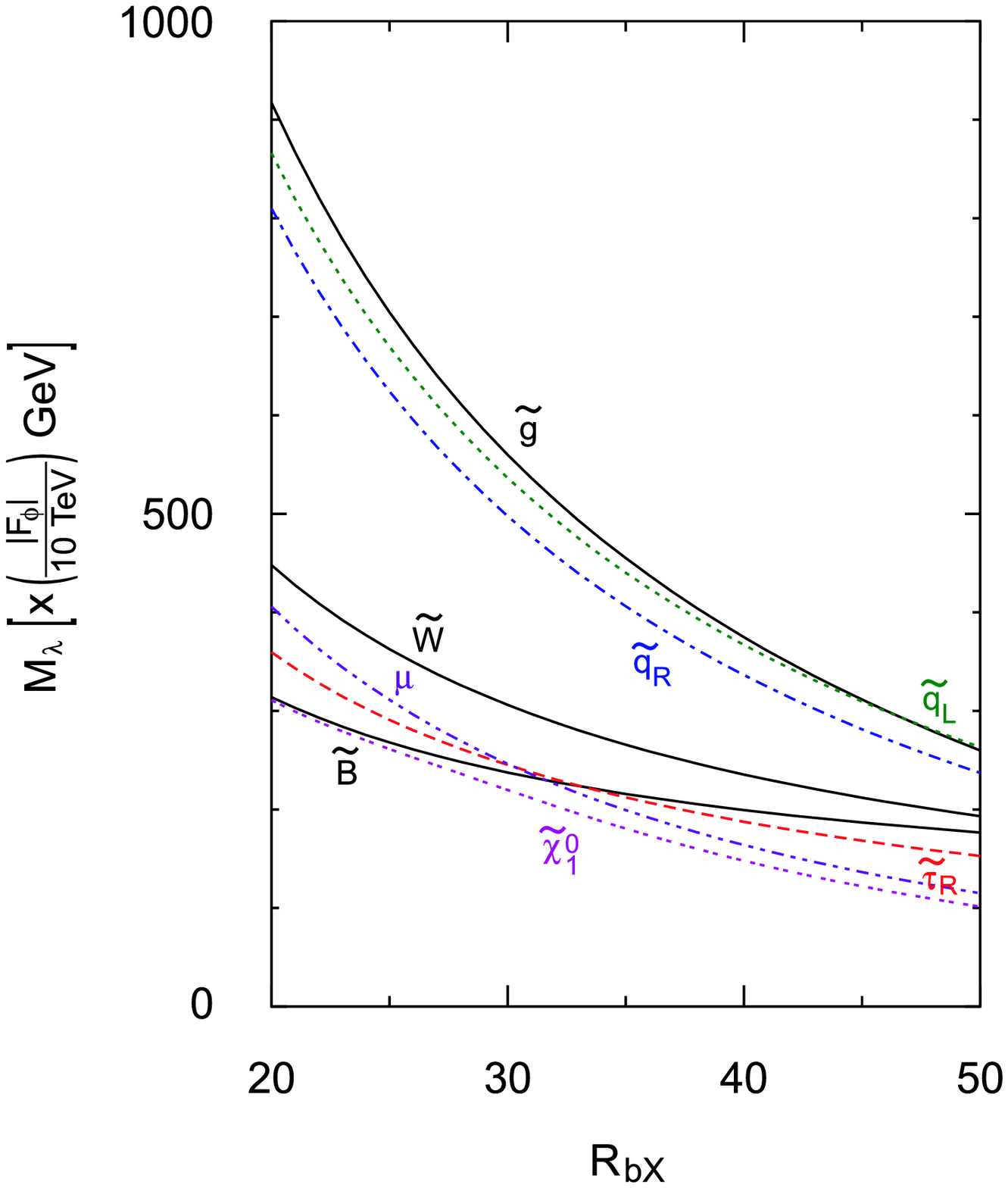}
\end{center}
\caption{Typical sparticle masses at the electroweak scale 
for $n=1$ moduli. We assume the universal K\"ahler terms for visible
and hidden sector fields; that is, $l={}^\forall m_i=1/3$ in
Eq.~(\ref{fXZQ}). The horizontal axis denotes the ratio of 
two $F$-term contributions: $R_{bX}\equiv F_\phi/(F_X/2X_R)$. In the
figure, $\tan\beta=10$, and $F_\phi$ is fixed to be 20~TeV as an
example, but the spectrum is scaled by $F_\phi$. The gaugino masses
are given by the solid (black) lines, and the two-dotted-dashed line
is the higgsino mass parameter $\mu$. We show the mass of the lightest
neutralino ($\tilde\chi^0_1$) by two-dotted line. The 
dotted (green) and dotted-dashed (blue) lines represent left- and
right-handed squarks of the first two generations, respectively. We
show, in particular, the mass of right-handed scalar tau by dashed
(red) line, which is the lightest among sfermions.}
\label{dilaton_mass}
\end{figure}

\begin{figure}[htbp]
\begin{center}
\includegraphics[scale=1]{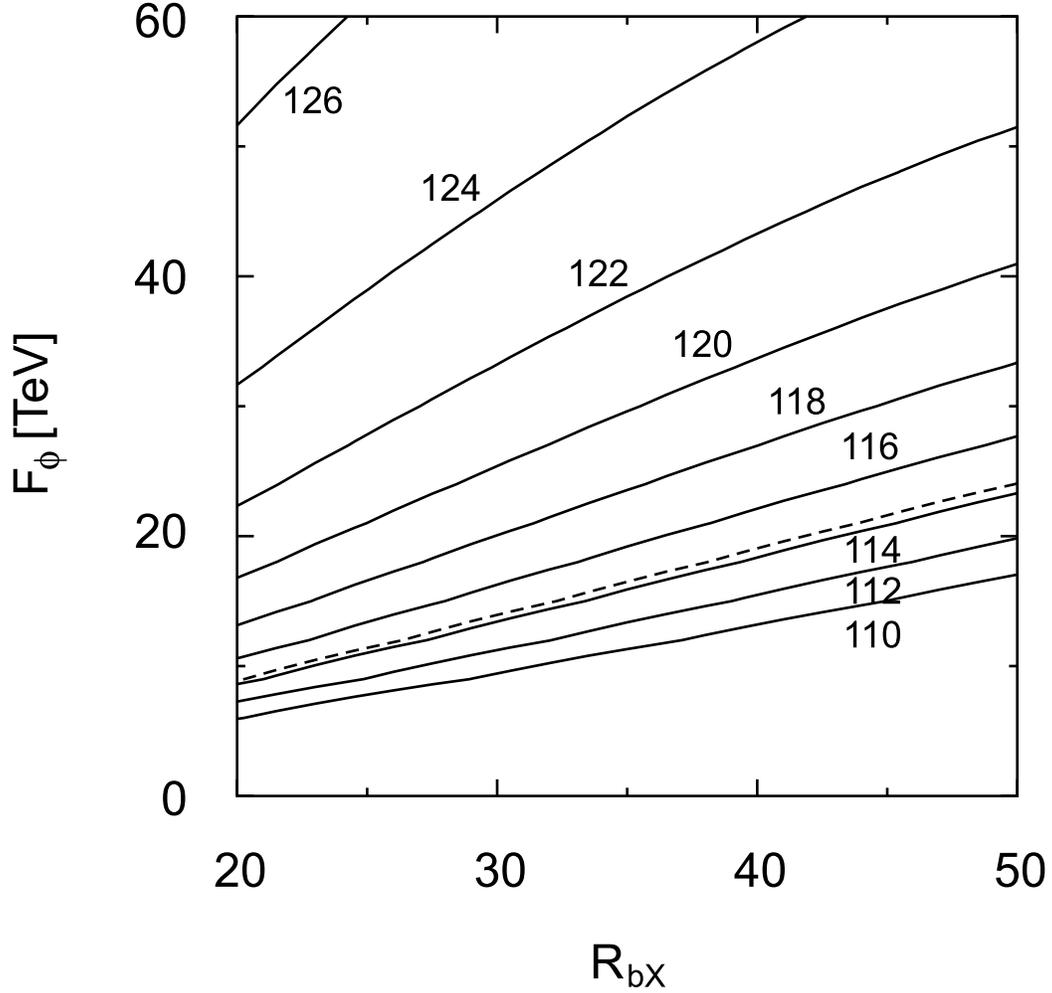}
\end{center}
\caption{Constant contours of the mass of the lightest neutral Higgs
boson for $n=1$ moduli. We assume the universal K\"ahler terms for
visible and hidden sector fields; that is, $l={}^\forall m_i=1/3$ in
Eq.~(\ref{fXZQ}). The dotted-dashed line denotes the current
experimental bound from LEP II\@. In the figure, we 
take $\tan\beta=10$ and the top quark pole mass $M_t=178$~GeV.}
\label{dilaton_Higgs}
\end{figure}

Let us examine the mass spectrum in some details for typical forms of
K\"ahler potentials. The first case 
is $n=1$ and $l={}^\forall m_i=1/3$. In this case, the moduli
contribution has a resemblance to the dilaton dominated scenario in
string-inspired supergravity. In Fig.~\ref{dilaton_mass}, we show the
mass spectrum for typical superparticles as the functions 
of $R_{bX}$. One can see from the figure that the mass hierarchy
among the superparticles is more suppressed for a larger value
of $R_{bX}$, i.e.\ with a larger suppression of the moduli
contribution for a fixed $F_\phi$. Here we have 
taken $F_\phi=20$~TeV\@. It is interesting that a lower bound of
sparticle masses is obtained by examining the experimental bound on
the lightest Higgs boson mass. We show in Fig.~\ref{dilaton_Higgs} the
mass contours of $M_h$ in the ($R_{bX},F_\phi$) plane. In the
numerical estimation of the Higgs mass, we used 
the {\tt FeynHiggs} package~\cite{FeynHiggs}. The requirement of a
heavy scalar top leads to a lower bound of the supersymmetry-breaking
scale $F_\phi$. It is also found from the figure that, when the
couplings in the moduli superpotential take natural 
values $a\sim O(1)$ and $c\sim$ TeV/Planck in the Planck 
unit, $R_{bX}$ is given by $\frac{2}{3}bX_R\sim 23$ and 
therefore $F_\phi$ must be larger than 10~TeV\@. This bound in turn
implies that the sparticles must be heavier than about 300~GeV (see
Fig.~\ref{dilaton_mass}). Furthermore the lightest superparticle is
the bino-dominant neutralino, but contains significant amount of the
higgsino component. This fact, however, relies on several assumptions,
for example, a choice of the index $l$ for hidden-sector field through
the $F$-term relation (\ref{FXFphi}) in the vacuum, and the higgsino
may become the lightest for other region. Thus, a general message is
that the lightest sparticle is given by a considerable mixture of
neutral gauginos and higgsinos. Finally, we note that the lower bound
of $F$ terms derived from the lightest Higgs mass bound is not very
sensitive to the value of $\tan\beta$ and $M_t$. In our numerical
analysis, we have taken $\tan\beta=10$ and $M_t=178$~GeV\@. When
$\tan\beta$ and/or $M_t$ increases, the Higgs boson mass receives
larger corrections, so the bound on $F$ imposed by the Higgs mass is
weakened. For instance, for $\tan\beta=40$ we numerically checked that
the Higgs mass increases only by about $1$--$2$~GeV, and thus the
bound on $F_\phi$ does not change significantly.

\begin{figure}[htbp]
\begin{center}
\includegraphics[scale=0.95]{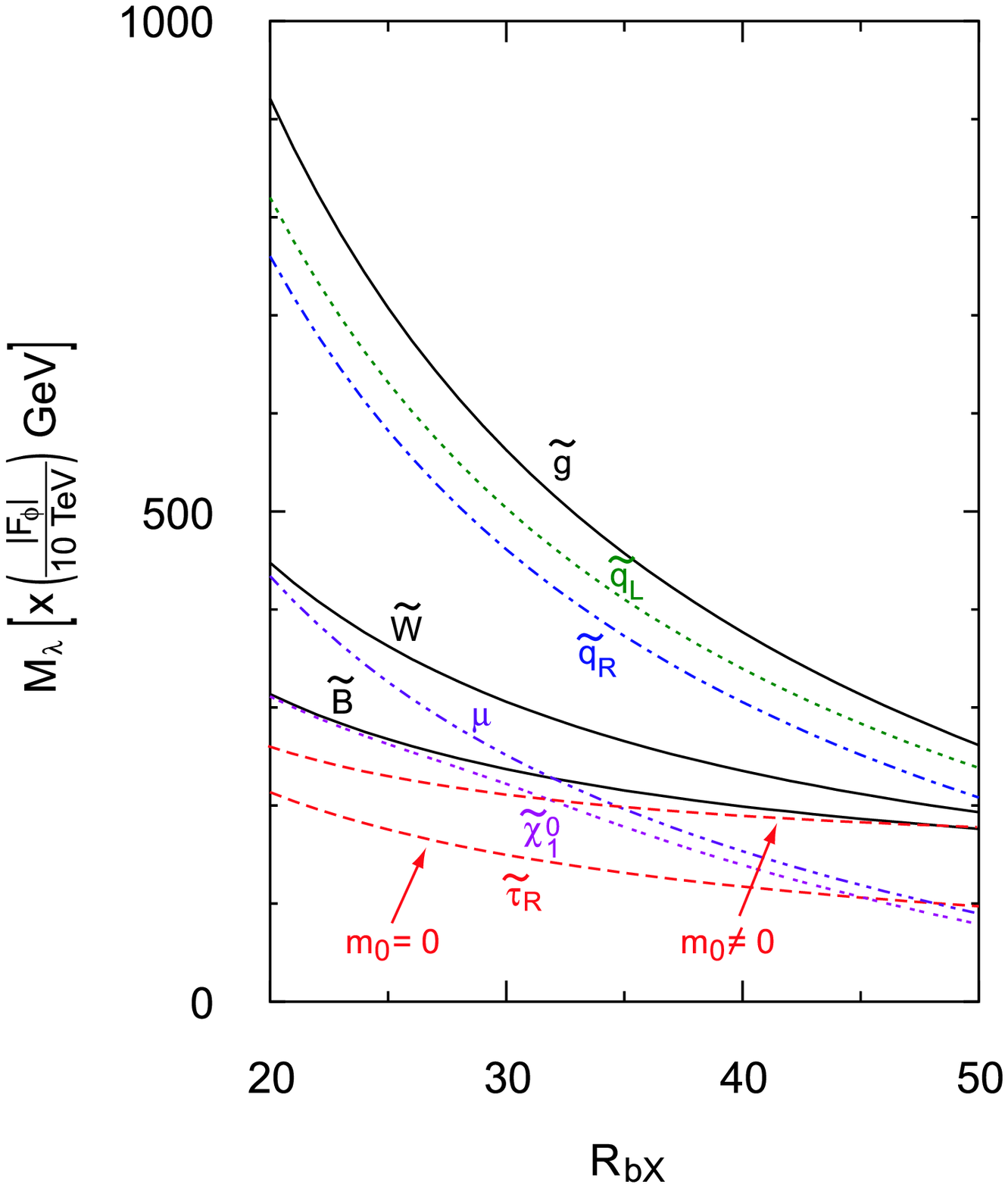}
\end{center}
\caption{Typical sparticle masses at the electroweak scale 
for $n=3$ moduli. We assume the sequestered K\"ahler potential for
visible and hidden sector fields; that is, $l={}^\forall m_i=0$ in
Eq.~(\ref{fXZQ}). The horizontal axis denotes the ratio of two $F$-term
contributions: $R_{bX}\equiv F_\phi/(F_X/2X_R)$. In the 
figure, $\tan\beta=10$, and $F_\phi$ is fixed to be 20~TeV as an
example, but the spectrum is scaled by $F_\phi$. The gaugino masses
are given by the solid (black) lines, and the two-dotted-dashed line
is the higgsino mass parameter $\mu$. We show the mass of the lightest 
neutralino ($\tilde\chi^0_1$) by two-dotted line. The dotted (green)
and dotted-dashed (blue) lines represent left- and right-handed squarks
of the first two generations, respectively. We show, in particular,
the mass of right-handed scalar tau by dashed (red) lines, which is
the lightest supersymmetric particle in a wide range of parameter
space for vanishing universal corrections to scalar 
masses ($m_0=0$). Also shown is the mass of right-handed scalar tau 
for $m_0=1.5 \times 10^{-2} |F_{\phi}|$.}
\label{moduli_mass}
\end{figure}

\begin{figure}[htbp]
\begin{center}
\includegraphics[scale=1]{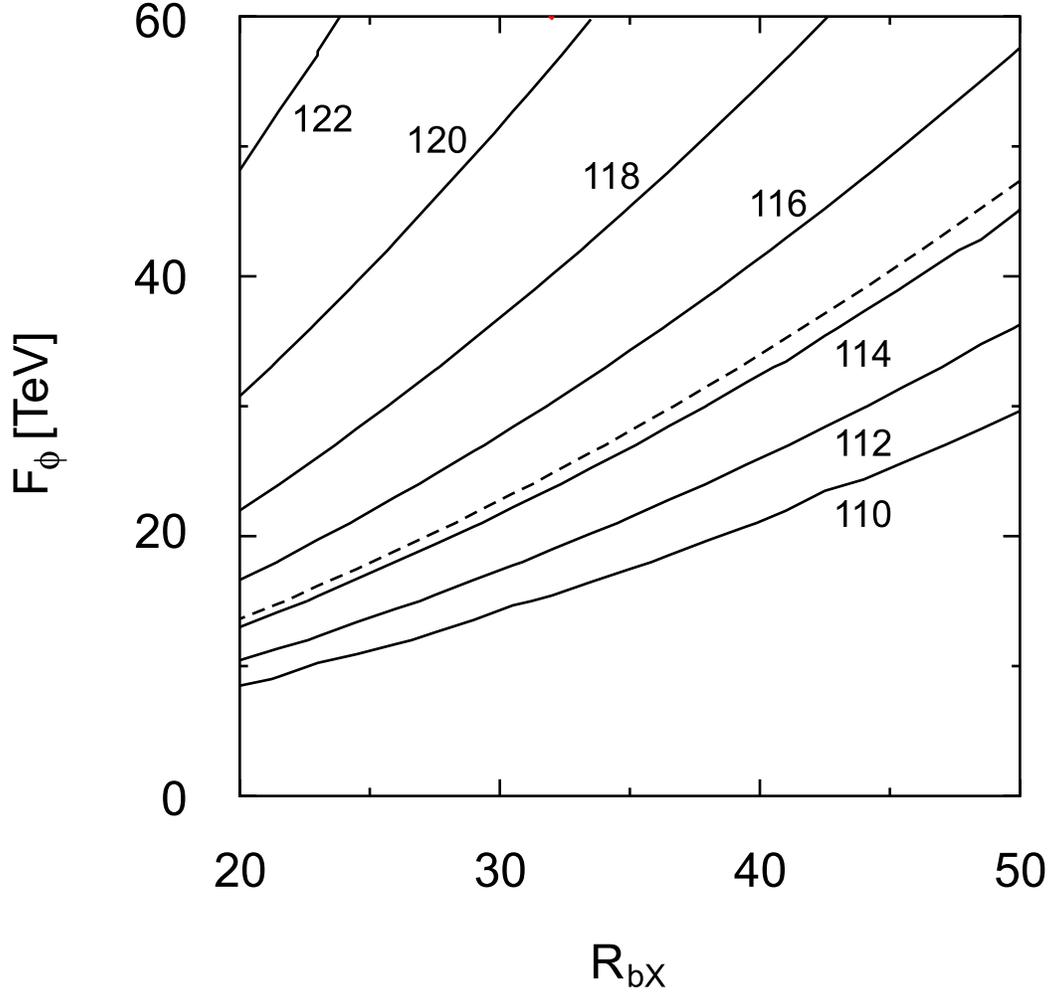}
\end{center} 
\caption{Constant contours of the mass of the lightest neutral Higgs
boson for $n=3$ moduli. We assume the sequestered K\"ahler potential
for visible and hidden sector fields; that is, $l={}^\forall m_i=0$ in
Eq.~(\ref{fXZQ}). The dotted-dashed line denotes the current experimental
bound from LEP II\@. In the figure, we take $\tan\beta=10$ and the top
quark pole mass $M_t=178$~GeV\@. We here include the 
universal $m_0$ corrections to sfermion masses to avoid the stable
scalar tau ($m_0=1.5\times 10^{-2} |F_\phi|$).}
\label{moduli_Higgs}
\end{figure}

The next example is for moduli with $n=3$, which often appear in
supergravity or superstring theory (e.g.\ K\"ahler size moduli in flux
compactifications). A simple choice of matter K\"ahler
potential is $l={}^\forall m_i=0$. Consequently, sfermions do not
receive soft masses from the moduli field. The spectrum at a
high-energy scale is therefore a class of the non-universal gaugino
masses (\ref{gaumass}) and sfermion masses given by the anomaly
mediation. It should be noted that the model does not suffer from the
problem of tachyonic sleptons in the pure anomaly mediation scenario
due to the heaviness of bino. That is, the gauginos obtain the
contribution from the light moduli field and lift up sfermion masses
via renormalization. In Fig.~\ref{moduli_mass}, we show the sparticle
mass spectrum at the electroweak scale. Again we have 
taken $F_\phi=20$~TeV\@. As in the case of $n=1$ moduli examined
above, we also find the lower bound of $F_\phi$ from the lightest
Higgs mass bound (Fig.~\ref{moduli_Higgs}). For natural values of the
superpotential couplings, $R_{bX}$ takes a value of about 35 and that
leads to a lower bound of $F_\phi\gtrsim 25$~TeV from the current
experimental result. Then the lightest neutralino mass has to be
larger than about 500~GeV\@. It is however noted that, in the present
simple case, the lightest superparticle is the scalar tau lepton in a
wide range of parameter space, which is slightly lighter than the
lightest neutralino.

\begin{figure}[htbp]
\begin{center}
\includegraphics[scale=1]{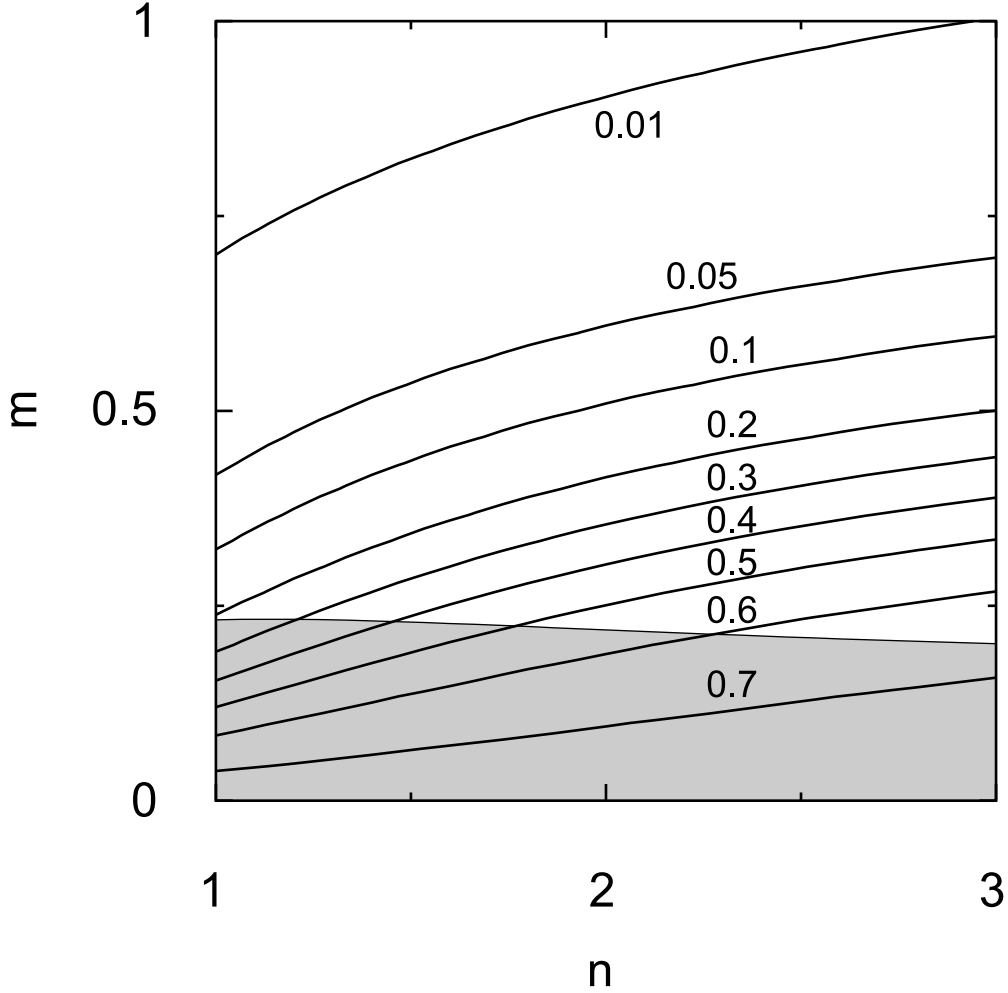}
\end{center} 
\caption{Constant contours of the higgsino composition $Z_h^2$ in the
lightest neutralino. Here $m$ denotes the universal violation of
sequestering between moduli and matter 
fields; $l={}^\forall m_i\equiv m$. Also shown is the shadow region in
which the scalar tau is lighter than  the lightest neutralino, and
thus cosmologically disfavored. In the 
figure, $\tan\beta=10$ and $F_\phi=20$~TeV are assumed. A larger value
of $F_\phi$ more increase the higgsino composition.}
\label{nm_higgsino}
\end{figure}

The existence of such a stable charged particle is unfavorable for
cosmology but does not necessarily mean a disaster. One could easily
incorporate some remedies to modify sfermion masses by introducing the
renormalization-group running above the gauge coupling unification
scale~\cite{RGaboveGUT} or some origins of additional contributions to
scalar masses. A possible source of the latter is given by weak
violation of the sequestering of the visible sector from the moduli
field in the K\"ahler potential. The sequestering is generically not
protected by some symmetry but rather may receive radiative
corrections. As a result, one could have some additional terms to soft
scalar masses. For example, consider the K\"ahler potential 
$K=-3\ln(X+\bar X-|Q|^2)-\ln(Y+\bar Y+\alpha(X+\bar X))+\cdots$ with 
tiny $\alpha$, which induces additional terms not only in scalar soft
masses but also in trilinear couplings. These two sources could be
parameterized by turning on the matter indices $m_i$ in the K\"ahler
potential. Thus we can systematically take into account
model-dependent scalar masses without affecting the gaugino sector for
a fixed value of $n$. In that case, $m_i$ are no longer integers, in
general. Similarly one may consider a correction to $n$. The result of
a numerical computation is depicted in Fig.~\ref{nm_higgsino} on 
the ($n,m$) plane, where we take the universal K\"ahler 
index, ${}^\forall m_i\equiv m$, for simplicity. In the shadowed
region, the scalar tau is lighter than the lightest 
neutralino. As $m$ increases, the neutralino becomes lighter than the
scalar tau. For instance, for $n=3$ moduli, a relatively small
correction, i.e.\ a small size of $m\gtrsim 0.2$, is sufficient to
make the scalar tau heavy and the theory becomes phenomenologically
viable. In the same figure, we also draw the contours of the higgsino 
composition $Z_h^2$ in the lightest neutralino, which is defined 
as $Z_h^2\equiv Z_3^2+Z_4^2$ where the lightest 
neutralino $\tilde\chi_1^0$ consists of 
$\tilde\chi_1^0=Z_1\tilde B+Z_2\tilde W+Z_3\tilde H_d+Z_4\tilde H_u$. 
When $Z_h^2$ is larger than 0.5, it is called higgsino-dominant. In
the figure, $F_\phi$ is fixed at 20~TeV, and $l=m$ has been assumed
for simplicity. For this choice of $F_\phi$ and $l$, there is the
higgsino-dominant region for small values of $m$, and the higgsino 
composition is still quite sizable, much larger than, say, 0.1 for
a wider region of the parameter space. It is
noticed that the higgsino composition becomes smaller 
as $m$ ($=l$ in this case) increases. This is because, for
larger $l$, $R_{bX}$ becomes smaller, increasing the mass difference
between the bino and the gluino. Thus the $\mu$ parameter,
which is controlled by the gluino mass, tends to be large relative to
the bino mass, making the higgsino composition small. Different
choices of $l$ may give slightly different results. In fact, we
considered the case where $l=0$ is fixed, and found the increase of
the higgsino composition, e.g.\ even a higgsino-dominant region
appears for smaller $m$, while the neutralino mass is kept smaller
than the scalar tau mass. Another thing we find is that if one
increases the scale $F_\phi$, the higgsino composition tends to
increase as well. That is understood from the fact that, for a 
larger $F_\phi$, the bino mass increases but the $\mu$ parameter
becomes smaller [see Fig.~\ref{gaugino_GUT} and
Eq.~(\ref{mu-fitting})].

Another simple way to cure the cosmological problem of sequestered
K\"ahler potential with $n=3$ moduli is to introduce additional
universal masses to sfermions. Such terms may arise in the K\"ahler 
potential due to the weak but non-negligible direct couplings between 
the visible and hidden sectors. Since we  assume a suppressed vacuum 
expectation value of the scalar component of the hidden sector field,
they do not contribute to the trilinear couplings. The magnitudes and
patterns of additional masses generally depend on the details of
models, for example, the origins of corrections, the configurations of
branes, and so on. But to incorporate such additional scalar masses
is a reasonable way to overcome the trouble with  too light sfermions,
often discussed in the literature. In the following, we assume
universal corrections to scalar masses, $m_0^2$, for simplicity. Then
the contributions are shown as:
\begin{equation}
  {m_Q^2}_i \,\longrightarrow\, {m_Q^2}_i + m_0^2.
\end{equation}
With a positive $m_0^2$, the mass of right-handed scalar tau is lifted
up. On the contrary, we notice that the masses of bino and higgsino do
not receive large corrections from $m_0^2$. In 
particular, $\mu$ depends weakly on the universal shift of the scalar
masses due to the cancellation among them. Actually, in the fitting
formula of $|\mu|^2$ (\ref{mu-fitting}), one can see that the
universal corrections from the scalar top and up-type Higgs cancel
out with each other, and $\mu$ is mainly controlled by the gluino
mass only. In the end, the property of the neutralino sector is
unchanged and the lightest neutralino, being an admixture of the bino
and the higgsino, becomes the lightest superparticle. For 
illustration, we add in Fig.~\ref{moduli_mass} the dashed red line
which depicts the mass of right-handed scalar tau for $m_0\neq 0$. Here
we have taken $m_0=1.5\times 10^{-2}|F_{\phi}|$ as an example. It is
numerically checked that the right-handed scalar tau becomes heavier
than the lightest neutralino if $m_0$ is larger 
than $1.2\times 10^{-2}|F_{\phi}|$ ($1.7\times 10^{-2}|F_{\phi}|$) 
for $\tan\beta=10$ (40), respectively. The dependence 
on $\tan\beta$ can be understood from the fact that the tau scalar
mass squared receives significant and negative radiative corrections
from $\tau$ fermion loops in the renormalization-group evolution with
large $\tau$ Yukawa coupling. We conclude 
that $m_0$ with ${\cal O}(10^{-2}|F_\phi|)$ can easily give the mass
spectrum with the neutralino lightest superparticle.

Finally, we comment on the case of strong violation of the sequestered
form of the K\"ahler potential. In particular, the absence of the
fourth-powered term $ZZ^\dagger QQ^\dagger$ is due to the assumption
of sequestering in the above analysis. This assumption requires a
special form of K\"ahler potential, that is, on a choice 
of $n$ and $m_i$. On the other hand, more general sets 
of $n$ and $m_i$ often lead to un-suppressed fourth-powered
terms. Then with non-vanishing $F$ terms of hidden-sector fields, they
induce soft masses for visible scalar particles. This contribution
could be involved in the analysis by introducing scalar 
masses $m_0$ again. If the K\"ahler potential has no sequestering form
at all, $m_0$ is expected to be comparable to $|F_\phi|$, when
considering the vanishing cosmological constant. Such a 
large $m_0$ exceeds contributions to the scalar masses from the moduli
and anomaly mediations by 1--2 orders of magnitude. We find from
numerical evaluation that the introduction of such large $m_0$ is
disfavored because of the failure of the electroweak symmetry
breaking. Too large additional scalar masses make the up-type Higgs
mass huge so that the conditions for radiative symmetry breaking are
not satisfied at least with the universal scalar masses at the
unification scale. On the other hand, it is found that the Higgs
bosons develop non-vanishing expectation values  
for $m_0/|F_\phi|\lesssim 10^{-1}$. Below that scale, the
neutralino may be light to the level of ${\cal O}(100)$~GeV, depending
on the value of $|F_\phi|$. As in the previous cases, the higgsino
mass tends to degenerate with that of the bino. Finally, we comment on
the effects of scalar trilinear couplings. In the setup above, we
neglected additional contributions to trilinear couplings. Such
contributions highly depend on the models and need more detailed
analysis with an explicit form of Lagrangian. Anyway the strong
violation of sequestering gives rise to a possibility of 
large $m_0$ comparable to $F_\phi$, which requires fine tuning among
scalar masses for successful electroweak symmetry breaking.

%%%%%%%%%%%%%%%%%%%%%%%%%%%%%%%%%%%%%%%%%%%%%%%%%%%%%%%%%%%%%%%%%%%%%%
\section{Cosmological Implications}
%%%%%%%%%%%%%%%%%%%%%%%%%%%%%%%%%%%%%%%%%%%%%%%%%%%%%%%%%%%%%%%%%%%%%%
\subsection{Gravitino and moduli problems}

Both gravitino and moduli are heavy with a small mass hierarchy
between them. The gravitino is heavier than supersymmetry-breaking
mass scale (squarks, etc.), and the moduli is still heavier than the
gravitino. Cosmological features with this mass hierarchy have
recently been studied in Refs.~\cite{KYY1,KYY2}. The heavy gravitino
decays before the big-bang nucleosynthesis starts. The hadronic
showers produced at the gravitino decay were shown to affect the
neutron to proton number ratio, changing thus the $^4$He
abundance~\cite{Kohri,hadronic}. Though the comparison with the
observations of $^4$He abundance yields the constraint on the
gravitino abundance, it is not as severe as the constraint from other
light elements. Thus considering the heavy gravitino scenario greatly
relaxes the otherwise severe gravitino problem. The moduli fields with
mass of order ${\cal O}(10^3)$~TeV or even higher can decay much
faster than 1 sec, and thus the reheating temperature after the moduli
decay is much higher than 1~MeV, which is compatible with the success
of the standard big-bang nucleosynthesis scenario~\cite{heavy-m,MYY,RT}.

In the scenario we presented, the moduli decay mainly into the gauge
bosons, but also decay to gravitinos with non-negligible branching
ratio. It was shown~\cite{KYY1,KYY2}, however, that the gravitinos
produced in this way do not spoil the big-bang nucleosynthesis in
certain regions of parameter space. Thus, we conclude that in our
scenario both the gravitino and the moduli problems can be solved
simultaneously.

%%%%%%%%%%%%%%%%%%%%%%%%%%%%%%%%%%%%%%%%%%%%%%%%%%%%%%%%%%%%%%%%%%%%%%
\subsection{Neutralino dark matter}

One of the fascinating features of supersymmetric models is that
the lightest superparticle is stable under the usual assumption
of $R$ parity conservation and thus it is a natural candidate for
dark matter in the Universe. Particular attention has been paid
to the case where the lightest among the neutralinos becomes the
lightest of the whole superparticles and thus a dark matter candidate.
In our present setup, since the gravitino and the modulinos, the
fermionic components of moduli supermultiplets, are rather heavy, the
lightest neutralino is likely the lightest among the superparticles in
the minimal supersymmetric standard model. Assuming the neutralino is
the lightest sparticle, we will briefly discuss this issue of
neutralino dark matter.

Let us first consider the relic abundance of the neutralino under the
assumption that the Universe underwent the standard thermal history,
namely, no entropy production nor non-thermal production of 
neutralinos takes place around and after the freeze-out of the
neutralinos. Then the computation of the thermal relic abundance can
be done in a standard matter. It should be compared with the value
inferred from observations. Inclusion of the WMAP data
implies~\cite{WMAP} the cold dark matter abundance to be
\begin{equation}
  \Omega_{\rm CDM}h^2 \,=\, 0.112^{+0.0161}_{-0.0181}, \quad
  (95~\%~{\rm C.L.})
  \label{omega}
\end{equation}
where $\Omega_{\rm CDM}$ is the density parameter of the cold dark
matter component, and $h$ is the Hubble parameter in units 
of 100 km/s/Mpc. The dark matter analysis was most extensively done in
the minimal supergravity scenario. It was 
shown~\cite{WMAP-SUSYDM} that the thermal relic abundance tends to be
too large, compared to the value obtained after the WMAP\@. To be
compatible with the observation, one needs some efficient annihilation
mechanisms, including (i) light neutralino and light 
sfermions, (ii) co-annihilation, (iii) resonance enhancement in Higgs
exchanges, and (iv) annihilation into W boson pair. Notice that the
last one is effective for wino or higgsino being the lightest
sparticle, the latter of which is realized only in the focus point
region~\cite{focus} in the minimal supergravity.

The situation in the present setup is a bit different from the minimal
supergravity case. As we emphasized before, the higgsino mass
parameter $\mu$ is relatively small due to the suppressed gluino
mass. Thus the lightest neutralino contains a significant amount of
higgsino components, which may enhance the annihilation cross
section. Though the thermal relic abundance of order 0.1 or so in
terms of the density parameter is expected, the actual value is,
however, rather sensitive to the portion of the higgsino components
and also to the mass of the lightest neutralino.

\begin{figure}[htbp]
\begin{center}
\includegraphics[scale=0.95]{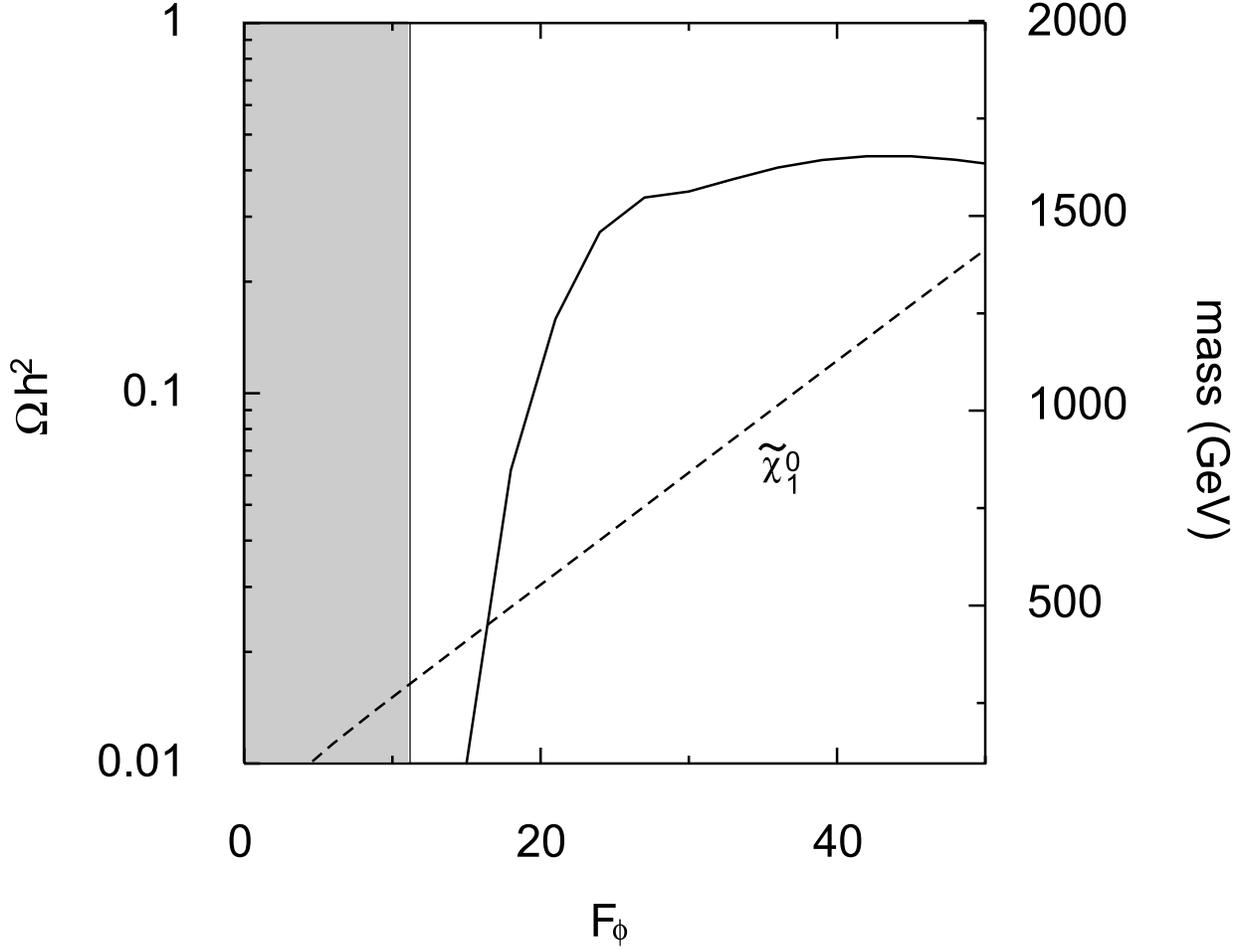}
\end{center}
\caption{Thermal relic density of the lightest supersymmetric particle
for $n=1$ moduli and $l={}^\forall m_i=1/3$. We take as an 
example $\tan\beta=10$ and $bX_R=35$. In the figure, we also plot the
mass of the lightest neutralino. The shadow region in the left is
excluded from the current experimental mass bound on the lightest
Higgs boson.}
\label{dilaton_relic}
\end{figure}

We now would like to quantify the argument given above. For this
purpose, we will illustrate the two cases:
(1) $n=1$, $\;l={}^\forall m_i=1/3$ and 
(2) $n=3$, $\;l={}^\forall m_i=0$ with extra universal contribution to
scalar masses, both of which cases were examined in the previous
section. In the numerical computation, we use
the {\tt DarkSUSY} package~\cite{DarkSUSY}. First let us discuss the
case (1) $n=1$, $\;l={}^\forall m_i=1/3$. As was pointed out, the
neutralino in this case is bino-dominant, but with a significant
portion of higgsinos. We show its thermal relic abundance in
Fig.~\ref{dilaton_relic}. In the same figure, the mass of the lightest
neutralino is also plotted. Here we have 
taken $bX_R=35$ and $\tan\beta=10$. Imposing the LEP-II bound on the
Higgs mass, only the region with $F_\phi\gtrsim 10$~TeV has been found
to be allowed (see Fig.~\ref{dilaton_Higgs}). On the other hand, it is
found from Fig.~\ref{dilaton_relic} that $F_\phi$ is roughly bounded
to be $F_\phi\lesssim 20$~TeV in order not to exceed the upper 
constraint (\ref{omega}) from the WMAP observation because of the
significant suppression of the neutralino relic abundance due to the
annihilation. Thus, for the 
range 10~TeV $\lesssim F_\phi\lesssim 20$~TeV, the thermal relic
density of the lightest superparticle is consistent with the WMAP
observation and also the Higgs boson mass bound is
satisfied. Moreover, the thermal relic becomes the dominant component
of dark matter on the upper side of the $F_\phi$ range. For a larger
value of $F_\phi$, to be consistent with the WMAP value, some dilution
mechanism should be in operation.
 
\begin{figure}[htbp]
\begin{center}
\includegraphics[scale=1]{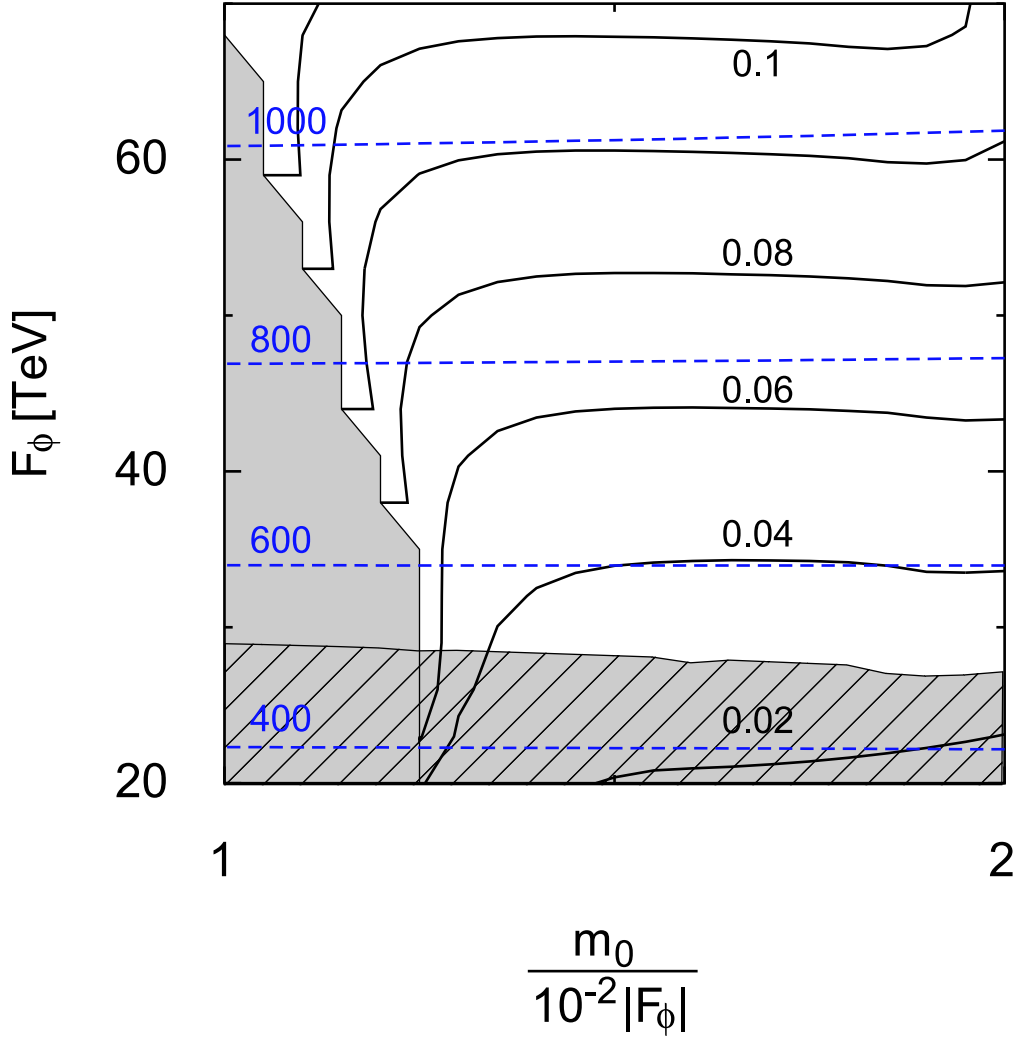}
\end{center}
\caption{Thermal relic density of the lightest supersymmetric particle
for $n=3$ moduli and $l={}^\forall m_i=0$. The solid lines are the
contours of the abundance, and the dashed blue ones denote the mass of
the lightest neutralino. We here introduce the universal scalar mass
correction $m_0$, and take as an 
example $\tan\beta=10$ and $bX_R=35$. The shaded region in the left is
excluded as the scalar tau being the lightest superparticle, and the
shadow region in the bottom is also experimentally disfavored from the
lightest Higgs boson mass.}
\label{moduli_relic}
\end{figure}

We then turn to the case (2) $n=3$, $\;l={}^\forall m_i=0$ with extra
universal contribution to scalar masses denoted 
by $m_0$. The $m_0$ contribution is added to make the scalar tau
heavier than the lightest neutralino, as was discussed in the previous
section. The lightest neutralino in this case is an (almost full)
admixture of higgsinos and bino (for the choice 
of $R_{bX}=35$). In fact, it is found that the ratio of the bino mass
parameter relative to the $\mu$ parameter varies 
from 0.9 to 1.1 as $F_\phi$ increases from 10 to 50~TeV\@. At the
same time, the higgsino component of the chargino degenerates with the
neutralino very well. We therefore expect efficient annihilation of
the neutralino as well as the neutralino-chargino coannihilation, and
hence relatively small relic abundance. Assuming the standard thermal
history, we obtain the numerical result 
in Fig.~\ref{moduli_relic}, where the contours of the 
relic abundance (solid lines) and the lightest neutralino mass (dashed
lines) are plotted. Here we have taken $\tan\beta=10$. In 
tiny $m_0$ region, the lightest sparticle becomes charged, namely, the
right-handed scalar tau, which should be excluded (shaded
region), and the shadow region in the bottom is also experimentally
disfavored from the lightest Higgs boson mass. We find from the figure
that, for a large portion of the parameter space, the relic abundance
of the neutralino is smaller than about 0.1, in accord with the WMAP
observation. Note, however, that it does not saturate the dark matter
density unless the lightest neutralino mass reaches 1~TeV region.

We have so far assumed the standard thermal history. Recently, it was
pointed out that the thermal history of the heavy gravitino/moduli
scenario may be very different from the standard
one~\cite{KYY1,KYY2}. The moduli decay release a huge entropy to
reheat the Universe, following the era where the moduli oscillation
dominates the energy density. The neutralinos in the primordial origin
will be completely diluted by the entropy production. The neutralinos
can, however, be produced in non-thermal origin. With the small mass
hierarchy among the moduli, gravitino, and superpartners of the
standard-model fields, the moduli can decay to gravitinos, followed by
the gravitino decay into lighter superparticles. They subsequently
decay to the lightest superparticles, i.e.\ the neutralinos. It was
shown that, in certain regions of the parameter space, the relic
abundance of the neutralino produced by this decay chain is in accord
with the dark matter density obtained from the WMAP observation, while
the decay products of gravitinos do not spoil the success of the
big-bang nucleosynthesis. Furthermore, the neutralinos may be
re-generated in the thermal bath if the reheating temperature is high
enough. The latter production may be important when the neutralino
contains a significant composition of higgsinos so that the
annihilation process is effective. It is a very interesting question
whether non-thermal mechanisms can yield the cold dark matter whose
abundance is in accord with the observations in our particular
setup. This issue will be discussed elsewhere.

Finally, we would like to make comments on direct/indirect detections
of the neutralino dark matter in the present scenario. The crucial
observation is, as we already repeated, the neutralino is an admixture
of the higgsino and the bino. Furthermore, as the gluino is relatively
light, the squark masses are also somewhat suppressed. As a result,
given a neutralino mass, the scattering cross section off the nuclei
as well as the annihilation cross section into photons, neutrinos, and
so on will be enhanced compared to, e.g.\ the minimal supergravity
case with the universal gaugino mass. A larger cross section increases
the chance of detection. On the other hand, in the setup we have
presented in this paper, the lightest neutralino tends to be
relatively heavy due to the Higgs boson mass bound. A heavier
neutralino then decreases the chance of detection. Some recent studies
on the effects of non-universal gaugino masses to the dark matter
detection (as well as the thermal relic abundance) can be found in
Ref.~\cite{nonuniv-DM}.

%%%%%%%%%%%%%%%%%%%%%%%%%%%%%%%%%%%%%%%%%%%%%%%%%%%%%%%%%%%%%%%%%%%%%%
\subsection{Potential destabilization and thermal effects}

In this subsection, we examine the global structure of moduli
potential in some detail. The global structure may be important when
one discusses inflation models in the framework of moduli
dynamics. Inflationary cosmology based on the KKLT compactification
has been discussed in the literature~\cite{inflation-KKLT}. Another
issue related to the global structure of moduli potential is the
overshooting problem~\cite{overshoot}. A scenario based on the KKLT
potential has been discussed to solve the
problem~\cite{overshoot-KKLT}. Here we would like to consider the
issue of potential destabilization due to the thermal effects at
finite temperature. Even though the moduli potential has a stable
minimum at zero temperature, the early stage of the Universe might be
hot enough to destabilize such a potential. In 
fact, Ref.~\cite{BHLR} discussed the thermal effects and consequent
destabilization of moduli potentials in various stabilization
mechanisms in superstring frameworks. Here we review their arguments
and would like apply them to the scheme we have presented.

The global structure of the moduli potential depends on the hidden
sector contributions as well as the form of K\"ahler potential. With
our assumption of K\"ahler potential (\ref{fW}) together with the
KKLT-type superpotential, we obtain the following scalar potential of
the moduli field:
\begin{equation}
  V \,=\, V(X_R) +V_H.
  \label{V}
\end{equation}
Here $V(X_R)$ denotes the potential which comes from the moduli part
(\ref{fX}) only and is explicitly written as
\begin{eqnarray}
  V(X_R) &\,=\,& \frac{1}{n(2X_R)^{n}} \bigg[\,
  \big[(2bX_R+n)^2-3n\big]|a|^2e^{-2bX_R} \nonumber \\
  && \qquad\qquad -2n(2bX_R+n-3)|ac|e^{-bX_R} +n(n-3)|c|^2\,\bigg]
  \label{VXR}
\end{eqnarray}
in the Einstein frame. On the other hand, the hidden sector
contribution $V_H$ has the form $V_H=(2X_R)^{l'}|W_Z|^2$, assuming the
perturbative expansion with respect to hidden-sector 
field $Z$. Here $l'=-2n/3-l$ from Eq.~(\ref{fXZQ}). From 
astronomical observations, the Universe is known to have almost flat
geometry, so the hidden-sector contribution should be introduced and
tuned to suppress the cosmological constant. We would like to stress
that the global structure of moduli potential is also affected by the
hidden-sector contribution $V_H$. Actually the whole 
potential $V$ generically has a barrier between the origin and
infinity in the moduli space, when considering vanishing
cosmological constant with negative $l'$. The formation of barrier
structure is understood as follows. Since $bX_R$ is positive in 
the present model, the second or third term in the bracket in
(\ref{VXR}) dominates the potential for large $X_R$, depending on a
choice of $n$ of the K\"ahler potential. Thus the potential goes to be
suppressed in the limit of large $X_R$. In addition, since $n$ is in
the range $0<n\leq3$, the potential has a negative value at its
minimum if one ignores the hidden sector contribution. Then taking into
account $V_H$ with negative $l'$ in order to tune the cosmological
constant, the potential around the minimum is uplifted so that the
barrier is obtained in the region between the true vacuum and the
infinity in the moduli space.

It is important to notice that the barrier structure may be washed out
by the thermal effects in the early Universe. The light particles,
which interact with each other, construct a thermal bath in the early
Universe. The effective potential at finite temperature acquires
additional terms which depend on both the temperature and coupling
constants. Since, in the present setup, a non-vanishing value of
moduli field contributes gauge coupling constants, the moduli 
value is affected by thermal effects. The free energy in
supersymmetric $SU(N_c)$ gauge theory~\cite{FT} is generally given by
\begin{equation}
  F \,=\, -\frac{\pi^2 T^4}{24}\Big[a_0+a_2g^2+O(g^3)\Big],
  \label{FE}
\end{equation}
where $g$ and $T$ are the gauge coupling and temperature,
respectively. The coefficient $a_0$ is $N_c^2+2N_cN_f-1$ with $N_f$
being the number of matter multiplets in the fundamental
representation, and the one-loop coefficient $a_2$ is
$\frac{-3}{8\pi^2}(N_c^2-1)(N_c+3N_f)$. The coupling dependence of the
free energy (\ref{FE}) leads to potential terms of moduli fields which
are related to the gauge coupling $g$~\cite{FTpot}. An important
point is that $a_2$ is negative. That is, a smaller value of gauge
coupling decreases the free energy. Since the gauge coupling 
constant ($1/g^2$) is shifted by $X_R$ in our scenario, the moduli
expectation value tends to be driven to infinity. This kind of
destabilization of the potential is rather different from other
cosmological problems such as the moduli or gravitino problem. If the
temperature becomes high enough in the early Universe, actually higher
than the so-called critical temperature, and once the moduli field
goes over its potential barrier, the moduli cannot be recovered to
take the original value even with high temperature due to late-time
entropy production. Thus the global structure of the potential is very
important to investigate the cosmological scenario.

The thermal effects have been studied for the KKLT models with $n=3$
moduli K\"ahler potential~\cite{BHLR}. Though in their analysis the
cosmological constant was tuned by introducing an anti-D3 brane, the
behavior is significantly unchanged to our case. The barrier arises at
low temperatures and disappears at the critical temperature
\begin{equation}
  T_{\rm crit} \,\simeq\, \sqrt{m_{3/2}}\,
  \bigg(\frac{3b}{4B}\bigg)^{\frac{1}{4}}g^{-\frac{3}{4}},
\end{equation}
where $B=(1/T^4)\partial F/\partial g(X_{\rm crit})$. The critical
temperature increases as the gravitino mass 
increases. For $m_{3/2}=10-100$~TeV, one 
finds $T_{\rm crit}\simeq 10^{11-12}$~GeV, which is $1$--$2$ orders of
magnitude larger than the case of $m_{3/2}=100$~GeV considered in
Ref.~\cite{BHLR}. Thus the reheating temperature is less constrained
in the heavy gravitino scenario.

\begin{figure}[htbp]
\begin{center}
\includegraphics[scale=0.65]{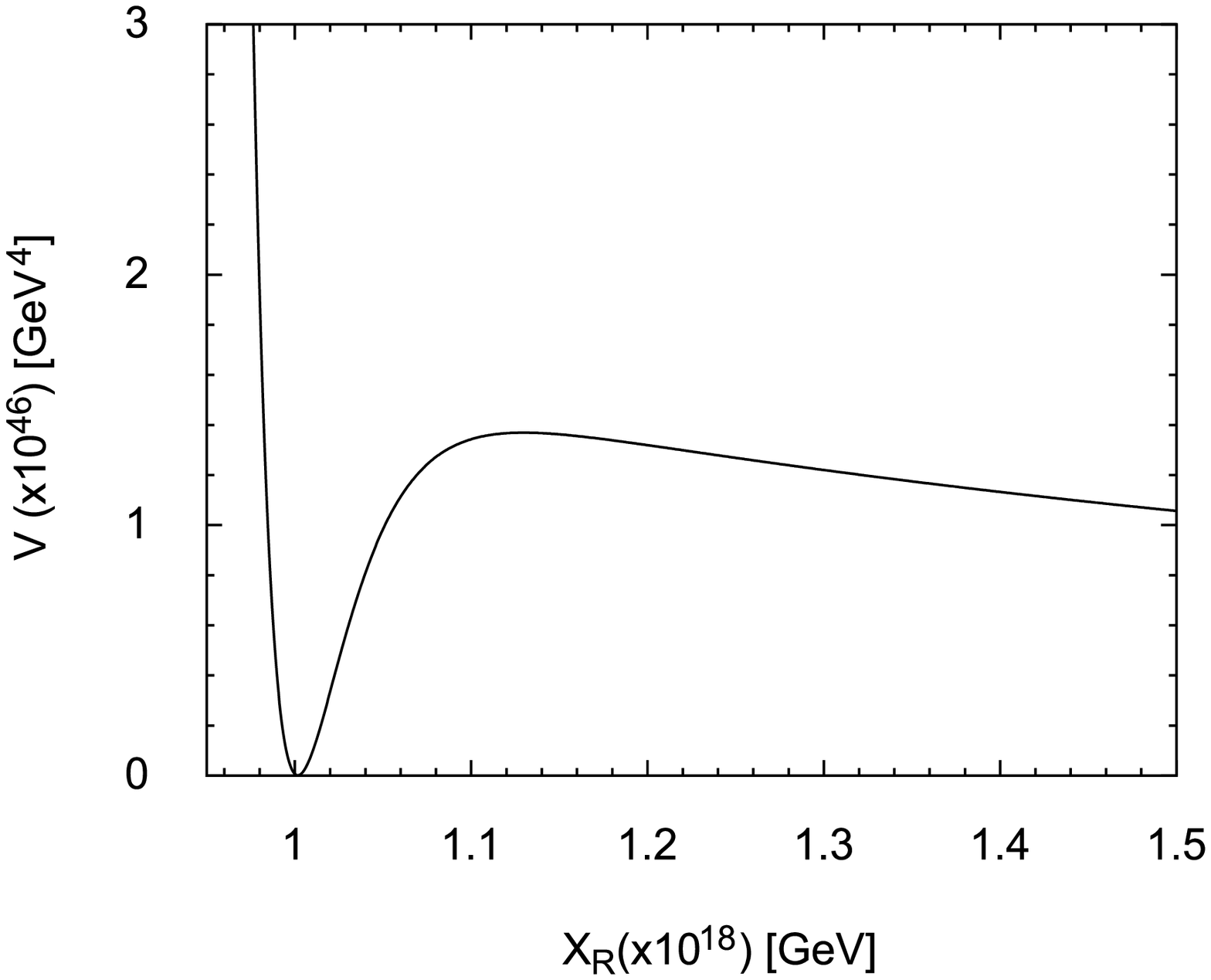}
\end{center}
\begin{center}
\includegraphics[scale=0.65]{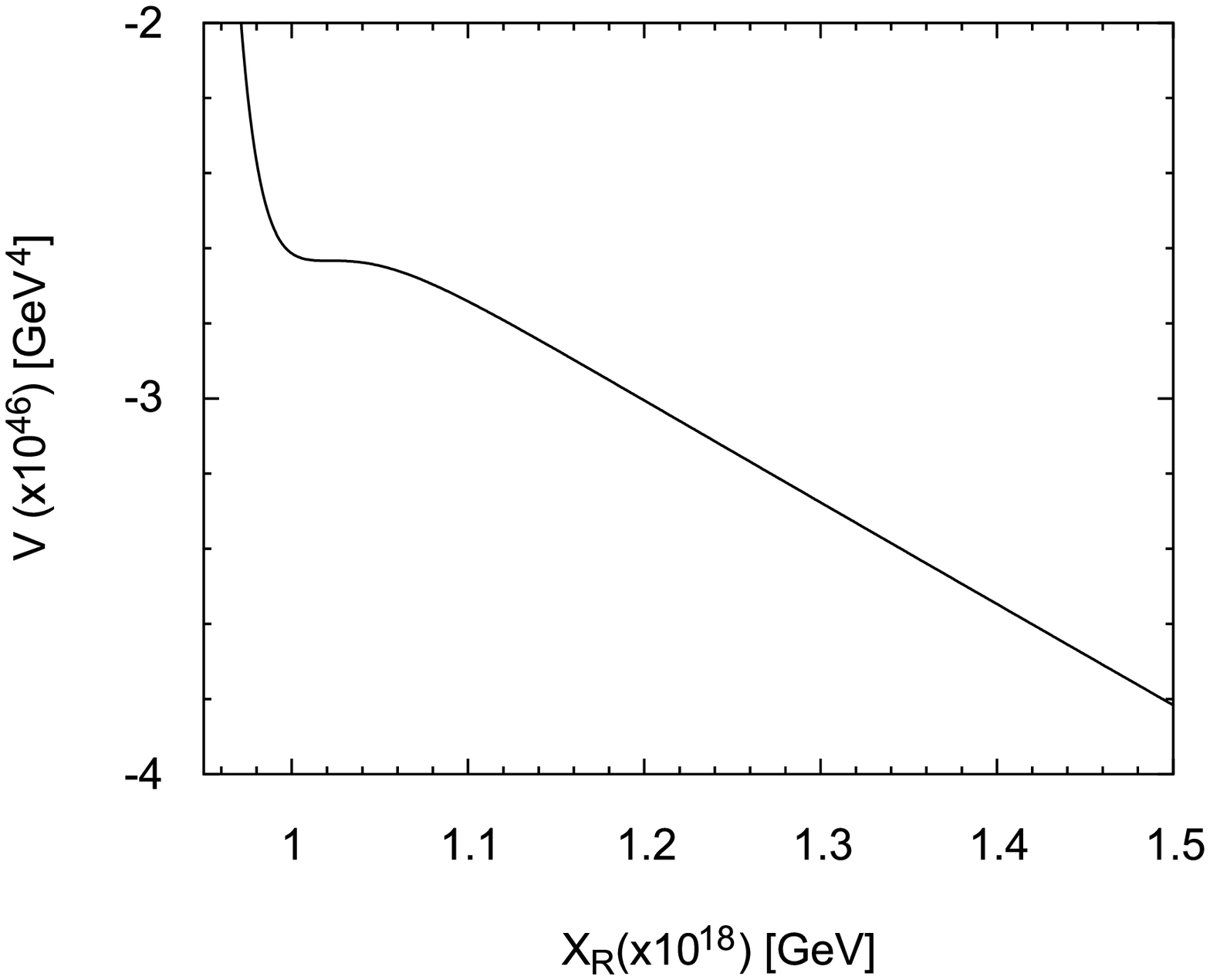}
\end{center}
\caption{The global structure of the moduli potential. Here we 
take $n=1$ and $l=1/3$. Above: the moduli potential at zero
temperature. Below: the same potential at the critical temperature.}
\label{dilaton_potential}
\end{figure}

A similar conclusion is drawn for $n=1$ moduli K\"ahler
potential. As an example, consider the sequestered form 
with $l'=-1$. By adjusting superpotential $|W_Z|^2$, the vacuum energy
is tuned to zero at the vacuum, and the barrier structure is formed in
the moduli potential again. Fig.~\ref{dilaton_potential} (top)
depicts the global structure of the potential at zero
temperature. Here the vacuum expectation value of the moduli is scaled 
to $X_R|_{\rm min}\simeq 1\times 10^{18}$~GeV\@. The thermal effects
might wash out this stable vacuum, and hence the moduli goes over the
barrier above the critical temperature. It can be easily confirmed
that, as it should be, $T_{\rm crit}\simeq 10^{11-12}$~GeV for the
gravitino mass scale in our scheme. Fig.~\ref{dilaton_potential}
(bottom) shows the moduli potential just at the critical
temperature. The scaling property 
of $T_{\rm crit}\sim\sqrt{m_{3/2}M_{\rm pl}}$ is rather a generic
result, which corresponds to the height of the potential barrier.

%%%%%%%%%%%%%%%%%%%%%%%%%%%%%%%%%%%%%%%%%%%%%%%%%%%%%%%%%%%%%%%%%%%%%%
\section{Summary}

We have investigated the moduli dynamics in the heavy gravitino
scenario. The constrained form of moduli superpotential has been
derived from a phenomenological requirement of CP conservation in 
gaugino masses. We have shown that the gaugino mass contributions from
the anomaly mediation and the moduli mediation are dynamically aligned
with each other and hence no CP phase arises from them, provided that
the superpotential has the special form of the sum over an 
exponential and a constant. Performing a K\"ahler transformation to
the supergravity Lagrangian, we can obtain another possible form of
moduli superpotential, namely, the one with two exponentials.

The superpotential and also the K\"ahler potential may be derived from
some underlying theory in high-energy regime. In fact, the former of
the two types of superpotential, dubbed the KKLT-type one, can be
obtained in superstring compactifications with fluxes, combined with
some non-perturbative effects. The fluxes may generate the constant
term of the superpotential, while the exponential term can be
attributed to non-perturbative dynamics, such as D-brane instantons
and gaugino condensation. The latter type of moduli superpotential is
realized by multiple (in this case, two) gaugino condensations. On the
other hand, our model-independent approach is based on the automatic
CP conservation in the soft terms, a highly phenomenological
argument. It is an interesting result that suppressing large CP
violation leads to the unique moduli potential which stabilizes the
moduli and triggers supersymmetry breaking.

The scalar potential for the moduli becomes rather steep around its
minimum due to the exponentials with large exponents, and thus the
moduli mass becomes much larger than the gravitino mass. In turn, this
implies the suppression of supersymmetry breaking in the moduli $F$
term. We have shown that with the KKLT-type superpotential the moduli
mediation and the anomaly mediation make comparable contributions in
size to the soft supersymmetry breaking masses. The resulting mass
spectrum of the superparticles in the minimal supersymmetric standard
model is very different from that of the minimal supergravity and of
the pure anomaly mediation. Therefore this scenario is testable in
future collider experiments by measuring the masses of
superparticles. One of the most important observations made in the
KKLT-type case is that the three gauginos belonging to different gauge
groups of the standard model tend to rather degenerate in mass at the
weak scale, which brings some characteristic features of the
superparticle mass spectrum. In particular, the lightest neutralino
becomes a considerable admixture of the higgsinos and the gauginos
(mostly bino). This should have some impact on collider signals as
well as neutralino dark matter searches.

We have briefly discussed cosmological implications of the
scenario. Heavy gravitino and moduli fields are cosmologically
welcome. They can solve the gravitino problem and also the moduli
problem associated with the huge late-time entropy production after the
moduli-dominated era. Much should remain to be seen in the cosmology
of this scenario. Particularly interesting and also important is to
establish a thorough cosmological evolution of this setup, including
inflation model and cosmological behavior of the moduli field.

It is also an interesting extension to include more numbers of moduli
fields which participate in supersymmetry breaking. That may be a
generic situation even after fluxes are introduced in superstring
compactifications. One should address the questions of CP phases in
gaugino masses as well as of induced flavor mixings and CP violation
in the sfermion sector. We leave such investigations to future work.

%%%%%%%%%%%%%%%%%%%%%%%%%%%%%%%%%%%%%%%%%%%%%%%%%%%%%%%%%%%%%%%%%%%%%%
\acknowledgments

The authors would like to thank K.~Choi for useful discussion. This
work was partially supported by a scientific grant from the Ministry
of Education, Science, Sports, and Culture of Japan (No.~14046201) and
by grant-in-aid for scientific research on priority areas: ``Progress
in elementary particle physics of the 21st century through discoveries
of Higgs boson and supersymmetry" (No.~16081202 and
No.~16081209). M.E.\ thanks the Japan Society for the Promotion of
Science for financial support.

\medskip

After completion of the work, we received a preprint~\cite{CFNO} which
considers related topics.

%%%%%%%%%%%%%%%%%%%%%%%%%%%%%%%%%%%%%%%%%%%%%%%%%%%%%%%%%%%%%%%%%%%%%%

%%%%%%%%%%%%%%%%%%%%%%%%%%%%%%%%%%%%%%%%%%%%%%%%%%%%%%%%%%%%%%%%%%%%%%
\newpage

\newpage

\newpage

\newpage

\newpage

\newpage

\newpage

\newpage

\newpage

\newpage

\end{document}